\documentclass[prb, letterpaper, twocolumn, fleqn, floatfix, showpacs, showkeys ]{revtex4}

\newcommand{\rcl}{R_\mathrm{cl}}

\newcommand{\ve}{\epsilon}

\newcommand{\ebg}{\epsilon_{\text{bg}}}

\newcommand{\bv}{{\bf v}}

\newcommand{\bR}{{\bf R}}
\newcommand{\bJ}{{\bf J}}

\newcommand{\br}{{\bf r}}
\newcommand{\bq}{{\bf q}}
\newcommand{\bQ}{{\bf Q}}

\newcommand{\bD}{{\bf D}}
\newcommand{\bE}{{\bf E}}

\newcommand{\bmu}{\mbox{\boldmath{\(\upmu\)}}}

\newcommand{\cA}{A}
\newcommand{\cF}{\mathcal{F}}
\newcommand{\cU}{\mathcal{U}}
\newcommand{\cN}{\mathcal{N}}
\newcommand{\cS}{\mathcal{S}}
\newcommand{\FcN}{\widetilde{\mathcal{N}}}
\newcommand{\DFcN}{\delta\!\widetilde{\mathcal{N}}}

\newcommand{\FG}{\widetilde{G}}
\newcommand{\FGo}{\widetilde{G}^{(0)}}

\newcommand{\nimp}{n_\mathrm{imp}}

\newcommand{\smin}{\sigma_\mathrm{min}}

\newcommand{\barrhozero}{\bar{\rho}_{0}}

\newcommand{\barn}{\bar{n}}

\newcommand{\Frho}{\widetilde{\rho}}
\newcommand{\Drho}{\delta\!\rho}
\newcommand{\DFrho}{\delta\!\widetilde{\rho}}
\newcommand{\SHD}{S_\mathrm{HD}}
\newcommand{\SSC}{S_\mathrm{SC}}

\newcommand{\DFphi}{\delta\!\widetilde{\phi}}
\newcommand{\DPhi}{\delta\!\Phi}
\newcommand{\Dphi}{\delta\!\phi}

\newcommand{\DFPhi}{\delta\!\widetilde{\Phi}}
\newcommand{\FD}{\widetilde{\Delta}}
\newcommand{\DN}{\delta\!N}

\usepackage{euscript, units, amsfonts, graphicx, dcolumn, fancyhdr}
\usepackage{times}
\usepackage{times, graphicx, units,mathrsfs,euscript,upgreek,amssymb,pifont}

\begin{document}

\title{Effects of the structure of charged impurities and dielectric environment on conductivity of graphene}

\author{R. Ani\v{c}i\'{c}}
\affiliation{Department of Applied Mathematics, University of Waterloo, Waterloo, Ontario, Canada N2L 3G1}
\author{Z. L. Mi\v{s}kovi\'{c}}
\email{zmiskovi@uwaterloo.ca}
 \affiliation{Department of Applied Mathematics, and Waterloo Institute for Nanotechnology, University of Waterloo, Waterloo, Ontario,
Canada N2L 3G1}

\date{\today}



\begin{abstract}

We investigate the conductivity of doped single-layer graphene in the semiclassical Boltzmann limit, as well as the conductivity minimum in neutral graphene within the self-consistent transport theory, pointing up the effects due to both the structure of charged impurities near graphene and the structure of the surrounding dielectrics.
Using the hard-disk model for a two-dimensional (2D) distribution of impurities allows us to investigate structures with large packing fractions, which are shown to give rise to both strong increase in the slope of conductivity at low charge carrier densities in graphene and a strongly sub-linear behavior of the conductivity at high charge carrier densities when the correlation distance between the impurities is large. On the other hand, we find that a super-linear dependence of the conductivity on charge carrier density in heavily doped graphene may arise from
increasing the distance of impurities from graphene or allowing their clustering into disk-like islands, whereas the existence of an electric dipole polarizability of impurities may give rise to an electron-hole asymmetry in the conductivity. Using the electrostatic Green's function for a three-layer structure of dielectrics, we show that finite thickness of a dielectric layer in the top gating configuration, as well as the existence of non-zero air gap(s) between graphene and the nearby dielectric(s) exert strong influences on the conductivity and its minimum. While a decrease in the dielectric thickness is shown to increase the conductivity in doped graphene and even gives rise to finite conductivity in neutral graphene for a 2D distribution of impurities, we find that an increase in the dielectric thickness gives rise to a super-linear behavior of the conductivity when impurities are homogeneously distributed throughout the dielectric.
Moreover, the dependence of graphene's mobility on its charge carrier density is surprisingly strongly affected, quantitatively and qualitatively, by the graphene-dielectric gap(s) when combined with the precise position of a 2D distribution of charged impurities.
Finally, we show that the conductivity minimum in neutral graphene is increased by increasing the correlation distance between the impurities, reduced by increasing the graphene-dielectric gap, and increased by decreasing the dielectric thickness in a top-gated configuration, even though the corresponding residual charge carrier density is reduced by decreasing the dielectric thickness.
\end{abstract}

\maketitle \thispagestyle{plain}

\begin{section}{Introduction}

Graphene is a realization of a two-dimensional (2D) material made of carbon atoms strongly bonded in a honeycomb--like lattice, exhibiting a Dirac-like spectrum for low-energy excitations of its $\pi$ electrons, which has been under intense scrutiny for possible applications in electronics, photonics,\cite{Avouris_2012}
and biochemical sensing.\cite{Allen_2010}
Being an all-surface material renders graphene extremely sensitive to
the incident electromagnetic fields and to the dielectric properties of the surrounding matter,\cite{Newaz_2012}
which is both a blessing and a curse from a technological point of view.
While the use of external gates and/or controlled adsorption of atomic and molecular species present an efficient means for inducing precise concentrations of charge carriers in graphene,\cite{Chen_2008}
the presence of indeterminate amounts of charged impurities, which may be trapped in a substrate or directly adsorbed on graphene, render quantitative details of many measurements of graphene's electronic and optical properties ''sample dependent''.\cite{Tan_2007}
In addition, integrating graphene in layered structures with different material properties may bring additional issues due to uncertainties in the geometric structure and the chemical composition of such structures.\cite{Fallahazad_2012,Hollander_2011}

Possibly the most intriguing manifestation of the presence of charged impurities is the famed minimum in the DC conductivity of single-layer graphene in the limit of vanishing doping, i.e., when the average density of induced charge carriers in graphene approaches zero.\cite{Tan_2007,Sarma_2011} It was shown that the minimum conductivity may be explained by the manifestation of a system of electron-hole puddles in graphene due to corrugation of the electrostatic potential that arises from a spatial distribution of the charged impurities in a substrate. \cite{PNAS_2007} On the other hand, the conductivity in heavily doped graphene layers often exhibits sub-linear behavior, or saturation with increasing charge carrier density, which is often explained by the presence of short-range scatterers in graphene, presumably arising from atomic-size defects in the carbon lattice. \cite{Yan_2011} However, it turned out that spatial correlation among the nearby charged impurities may provide an alternative and more plausible explanation of the conductivity saturation in single-layer graphene. \cite{Li_2011,Yan_2011} Moreover, the atoms adsorbed on graphene often show tendency of clustering and forming islands, which may additionally affect the mobility of charge carriers in graphene. \cite{McCreary_2010}

As far as the structure and composition of the surrounding material is concerned, preference is usually given to insulators and metals that only engage in weak interactions with graphene of the van der Waals type, leaving the structure of its $\pi$ electron bands largely intact in the vicinity of the Dirac point.\cite{Wehling_2009} Those interactions are characterized with relatively large spatial gaps between graphene and the nearby material, on the order of several \AA ngstr\"{o}ms, which reduce the dielectric screening by that material and often exhibit significant fluctuations in their size due to the surface roughness of the material.\cite{Ishigami_2007}
Furthermore, when graphene is top gated with a layer of high-$\kappa$ dielectric material, the mobility of its charge carriers may be affected by a strong image interaction with
the metallic top gate.\cite{Fallahazad_2012,Hollander_2011,Ong_2012}
Finally, for electrolytically top-gated graphene, the presence of mobile ions in the nearby electrolyte may provide additional screening of the charged impurities in a solid substrate.\cite{Chen_2009,Miskovic_2012}

All of the above examples of the effects of charged impurities near graphene and the structure of the surrounding dielectrics play
important roles in its charge carrier transport, plasmon dispersion in doped graphene, and graphene's capacitance, which are of interest in electronics, photonics, and sensing, respectively.
It was recently shown that those effects may be conveniently modeled by using Green's function (GF) for the Poisson equation for a layered structure,
\cite{Ong_2012,Miskovic_2012}
which is easily combined in a self-consistent manner with the polarization function of graphene within the random phase approximation (RPA) when graphene is modeled as a zero-thickness material.
\cite{Castro_2009}
In this work, we illustrate such approach to modeling the conductivity of single-layer graphene with large area by considering a three-layer structure of the surrounding dielectrics and using an expression for the conductivity that results from the semiclassical Boltzmann transport (SBT) theory for doped graphene.
\cite{Sarma_2011}
However, that expression is derived here via the Energy loss method (ELM),\cite{Gerlach_1986}
which explicitly evaluates the friction force on a system of external charges with the spatial distribution that moves rigidly parallel to graphene.\cite{Allison_2009,Allison_2010,Radovic_2012}
Hence, the ELM has an added utility as it may be used in studying other processes, such as sliding friction of molecular layers physisorbed on graphene,\cite{Krim_2012}
or probing the streaming potential in a flowing electrolyte by a graphene based sensor,\cite{Newaz_2012_b}
which will be tackled in future work.

In this work we focus on several effects in the DC conductivity of graphene.
First, we explore the effects of long correlation distances among impurities that give rise to large packing fractions, which cannot be described by a simple step-like pair correlation function.\cite{Yan_2011,Li_2011} For that purpose we use an analytically parameterized model of hard disks (HD) due to Rosenfeld,\cite{Rosenfeld_1990} which gives
reliable results for packing fractions up to the freezing point of a 2D fluid.
Next, whereas all the previous studies assumed that charged impurities reside in a plane parallel to graphene, our statistical formulation of the theory allows for a fully three-dimensional (3D) spatial distribution of impurities that may reside at a range of distances from graphene. In addition, we allow that individual impurities may be characterized by atomic-like form factors, which include a finite dipole moment and a spatial spread that accounts for the existence of disk-like clusters near graphene. Furthermore, by taking advantage of the electrostatic GF for a three-layer structure, we also study the effects that arise in conductivity of graphene due to finite thickness of a nearby dielectric and a finite gap between graphene and the nearby dielectrics. Finally, the above effects are also studied in the context of the conductivity minimum within the Self-consistent transport (SCT) theory.\cite{PNAS_2007}

Specifically, in this paper we show via the HD model that large correlation distances between charged impurities may give rise to significantly larger initial slopes of the conductivity (or larger mobility) at lower charge carrier densities, as well as to a more pronounced saturation, or sub-linear behavior of conductivity at higher densities than in the case of small correlation distances. Next, the effects of clustering of charge impurities, as well as the increasing distance from graphene are confirmed to give rise to super-linear dependence of conductivity on charge carrier density in heavily doped graphene, in agreement with observations\cite{McCreary_2010}
and modeling,\cite{PNAS_2007} respectively. Impurities with finite dipolar polarizability are shown to give rise to electron-hole asymmetry in the conductivity as the sign of charge carrier density changes, which may be related to experimental observations in some graphene samples.\cite{Tan_2007}
Regarding the geometrical factors of a nearby dielectric layer, we find an increase in both the conductivity and mobility of graphene when the layer thickness decreases in the case of a 2D distribution of impurities, whereas a homogeneous 3D distribution of impurities gives rise to a super-linear behavior of the conductivity with increasing layer thickness. Most intriguingly, we find a strong effect on the mobility of graphene due to the presence of a finite gap between graphene and the nearby dielectrics in conjunction with the varying position of impurities, which was not previously considered in the modeling of the transport properties of graphene, but was observed in studying the polarization forces on external charges.
\cite{Allison_2009}
Finally, a similarly strong effect of the finite gap between graphene and a nearby dielectric is also demonstrated in the minimum conductivity within the SCT theory.\cite{PNAS_2007}

After outlining the theoretical model in the next section, we discuss our numerical results, and give concluding remarks. In the
Appendices we outline a derivation of the electrostatic GF and provide details for several models of the charged impurity structure.
Note that, unless otherwise explicitly stated, we use gaussian electrostatic units where $4\pi\ve_0\equiv 1$, with $\ve_0$ being
the dielectric permittivity of vacuum.

\end{section}

\begin{section}{Theory}

We assume that a single-layer graphene sheet of large area is embedded into a stratified structure so that it lies parallel to layers of various dielectrics with abrupt interfaces among them, as shown in Fig.~1. Using a 3D Cartesian coordinate system with coordinates $\bR\equiv\{\br,z\}$, the entire structure may be then considered translationally invariant (and is assumed to be isotropic) in the directions of a 2D position vector $\br=\{x,y\}$.
Furthermore, assume that a system of charged particles is distributed throughout the structure and is moving rigidly at a constant velocity $\bv$ parallel to graphene. If the stationary volume density of charges in the moving frame of reference is given by $\rho_0(\bR)\equiv\rho_0(\br,z)$, then the corresponding volume density in the rest frame of graphene (the laboratory frame of reference) is given by $\rho(\bR,t)=\rho_0(\br-\bv t,z)$.
\begin{figure}
\centering
\includegraphics[width=0.4\textwidth]{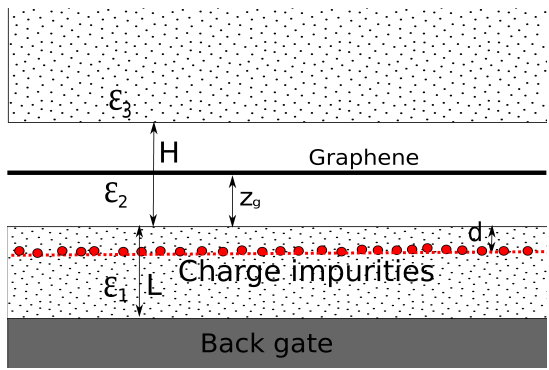}
\caption{(Color online)
Diagram showing a three-layer structure of dielectrics with the relative bulk dielectric constants $\ve_j$ for $j=1,2,3$, which occupy the regions defined
by the intervals $I_1=[-L,0]$, $I_2=[0,H]$ and $I_3=[H,\infty)$ for the $z$ coordinate of a Cartesian coordinate system, respectively.
}
\end{figure}

This notion of a rigidly moving distribution of external charges may be related to several realistic physical situations where the relative motion of particles
with respect to each other may be treated as adiabatic at the time scale of the charge carrier dynamics in graphene. Examples include sliding of a film of adsorbed molecular layers across graphene,\cite{Krim_2012} flow of a molecular fluid that contains dissolved ions in thermal equilibrium,\cite{Newaz_2012_b} or propagation of ionized fragments that result from planar Coulomb explosion of a cluster grazingly scattered from graphene. \cite{Song_2005} In each of those examples, the movement of external charged particles gives rise to energy dissipation due to excitations of charge carriers in graphene.

Conversely, one my reverse the frames of reference and consider the regime of steady-state electric conduction in graphene where its charge carriers move with a constant drift velocity $-\bv$. In this case the distribution of external particles is static in the laboratory frame and hence may be used to model fixed charged impurities near graphene. If the speed $v=\|\bv\|$ is sufficiently low, then the electrical resistivity of graphene may be related to energy dissipation due to scattering of its charge carriers on external charged impurities, giving rise to Ohmic heating of graphene.
This idea of reversing the frames of reference is a basis of the ELM that was developed for studying the transport properties of semiconductor heterostructures by means of the dielectric response formalism for their conducting electrons.\cite{Gerlach_1986}
This method was used successfully in studying the scattering of conduction electrons on interface roughness\cite{Kaser_1995}
and polarizable scattering centers,\cite{Kaser_1997}
as well as in discussing vibrational damping in adsorbed layers due to surface resistivity,\cite{Persson_1991}
and in studying optical properties of thin films for solar energy materials.\cite{Jin_1988}
Moreover, this same idea of the equivalence of a drag force on a uniformly moving system of impurities and the total force on the electron fluid in doped graphene was recently applied to evaluate the conductivity of graphene within the semiclassical hydrodynamic model for its charge carriers.\cite{Mendoza_2013}

We note that the ELM gives an expression for the conductivity of doped graphene, which is identical to that obtained by the SBT theory,\cite{Sarma_2011} but we chose ELM because it yields the drag force on externally moving charges as a side result that may be more directly used in modeling other processes, such as sliding friction of molecular layers physisorbed on graphene
\cite{Krim_2012}
or probing the streaming potential in a flowing electrolyte by a graphene based sensor,\cite{Newaz_2012_b}
to mention a few.

\subsection{Energy loss method}

To be specific, we assume that the system of external charges consists of $N$ particles, each carrying a total charge of $Z_je$ (where $e>0$ is the proton charge) that is distributed around the center of the particle according to some function $\Delta_j(\bR)$, such that $\int d^3\bR\,\Delta_j(\bR)=Z_j$ with $j=1,2,\ldots,N$.
If the $j$th particle is centered at the position  $\bR_j=\{\br_j,z_j\}$ in the moving frame of reference, we may write for the total density of charges in that frame
\begin{eqnarray}
\rho_0(\br,z)=e\sum_{j=1}^N\Delta_j\!\left(\br-\br_j,z-z_j\right).
\label{rhoCM}
\end{eqnarray}
Given that the positions $\bR_j$ of external particles, as well as their individual charge densities $\Delta_j(\bR)$ are statistically distributed, we denote their joint ensemble average by $\langle\cdots\rangle$. Assuming that this distribution is translationally invariant in the directions of $\br$, we note that $\langle\rho(\bR,t)\rangle=\langle\rho_0(\br,z)\rangle\equiv\barrhozero(z)$ can only be a function of the perpendicular coordinate $z$. Therefore, assuming that the
equilibrium areal number density of charge carriers is uniform across graphene, its value $\barn$ will be determined by both the function
$\barrhozero(z)$ and the potential applied through the external gates. We assume that $\barn$ has a sufficiently large value allowing us to neglect the effects of fluctuations in the charge carrier density in graphene on its screening properties. On the other hand, we assume $\barn$ to be small enough to allow the use of a 2D response function for graphene's $\pi$ electrons in the approximation of Dirac fermions.
\cite{Wunsch_2006,Hwang_2007}
Those requirements practically limit our considerations of graphene's DC conductivity within the SBT theory to an approximate range of doping densities 10$^{11}$ cm$^{-2} \lesssim \barn  \lesssim$ 10$^{13}$ cm$^{-2}$ (we assume $\barn>0$ unless stated otherwise).

We further define the fluctuation in the charge density of external particles by $\Drho(\bR,t)\equiv\rho(\bR,t)-\langle\rho(\bR,t)\rangle=\rho_0(\br-\bv t,z)-\langle\rho_0(\br,z)\rangle\equiv\Drho_0(\br-\bv t,z)$ and use it in the Poisson equation, allowing us to express the resulting fluctuation of the electrostatic potential, $\DPhi(\bR,t)$, in terms of the electrostatic GF for the entire system, $G(\bR,\bR';t-t')\equiv G(\br-\br';z,z';t-t')$, as
\begin{eqnarray}
\DPhi(\bR,t)=\int d^3\bR'\,\int\limits_{-\infty}^\infty dt'\,G(\bR,\bR';t-t')\,\Drho(\bR',t').
\label{DPhi}
\end{eqnarray}
Using a tilde to denote the Fourier transform (FT) of various quantities with respect to the 2D position ($\br\rightarrow\bq$) and time ($t\rightarrow\omega$), the above expression is recast in the form
\begin{eqnarray}
\DFPhi(\bq,z,\omega)=\int\limits_{-\infty}^\infty dz'\,\FG(\bq;z,z';\omega)\,\DFrho(\bq,z',\omega),
\label{DFPhi}
\end{eqnarray}
where
\begin{eqnarray}
\DFrho(\bq,z,\omega)&=&\int d^2\br\int\limits_{-\infty}^\infty dt\,\mbox{e}^{-i\bq\cdot\br+i\omega t}\,\Drho_0(\br-\bv t,z)
\nonumber\\
&=&2\pi\,\delta(\omega-\bq\cdot\bv)\,\DFrho_0(\bq,z)
\label{DFrho}
\end{eqnarray}
defines the relation between the FTs of the fluctuations of the external charge densities in the two reference frames.
Here, $\DFrho_0(\bq,z)=\Frho_0(\bq,z)-(2\pi)^2\,\delta(\bq)\,\barrhozero(z)$ is defined via the FT of the external charge density in the moving frame of reference,
\begin{eqnarray}
\Frho_0(\bq,z)=e\sum_{j=1}^N\FD_j(\bq,z-z_j)\,\mathrm{e}^{-i\bq\cdot\br_j}.
\label{DFrho_0}
\end{eqnarray}

It may be shown that the ensemble average of the energy loss rate is given by
 \cite{Mowbray_2010}
\begin{eqnarray}
\left\langle\frac{dW}{dt}\right\rangle&=&-\int d^3\bR\,\left\langle\Drho(\bR,t)\,\frac{\partial}{\partial t}\DPhi(\br,z,t)\right\rangle
\nonumber \\
&=& i\int\frac{d^2\bq}{\left(2\pi\right)^2}\,(\bq\!\cdot\!\bv)\,\int dz\int dz'\,\FG(\bq;z,z';\bq\!\cdot\!\bv)
\nonumber \\
&&\times
\left\langle\DFrho_0(-\bq,z)\DFrho_0(\bq,z')\right\rangle.
 \label{Eloss}
\end{eqnarray}
On using the symmetry properties of the FT of the full GF, $\FG(\bq;z,z';\omega)=\FG(-\bq;z',z;\omega)$ and $\FG^\mathrm{(cc)}(\bq;z,z';\omega)=\FG(-\bq;z,z';-\omega)$, where $\mathrm{(cc)}$ denotes complex conjugation, one notices that only the imaginary part of the factor $\FG(\bq;z,z';\bq\!\cdot\!\bv)$ in Eq.~(\ref{Eloss}) contributes to the energy loss. Furthermore, assuming that graphene has a zero thickness and is placed in the plane $z=z_g$, we may express $\FG(\bq;z,z';\omega)$ in terms of the (real valued) 2D FT of the GF $\FGo(\bq;z,z')$ for the dielectric environment \emph{without} graphene, as given in Eq.~(\ref{Dyson}).
Thus, Eq.~(\ref{Eloss}) may be rewritten as
\begin{eqnarray}
\left\langle\frac{dW}{dt}\right\rangle &=&\int\frac{d^2\bq}{\left(2\pi\right)^2}\,V_C(q)\,
(\bq\!\cdot\!\bv)\,\Im\!\left[\frac{-1}{\ve(q,\bq\!\cdot\!\bv)}\right]
\nonumber \\
&&\times
\left\langle\DFcN(-\bq)\DFcN(\bq)\right\rangle,
 \label{Eloss_ave}
\end{eqnarray}
where we have defined a dielectric function that describes the dynamic screening of external electrostatic fields in the plane $z=z_g$ due to the polarization of the entire system as
\begin{eqnarray}
\ve(q,\omega)=\ebg(q)+V_C(q)\,\chi(q,\omega),
 \label{eps}
\end{eqnarray}
with $\ebg(q)\equiv 2\pi/\left[q\FGo(q;z_g,z_g)\right]$ being an effective background dielectric function due to the polarization of the system \emph{without} graphene,
$V_C(q)=2\pi e^2/q$ the in-plane FT of the Coulomb potential, and $\chi(q,\omega)$ a 2D polarization function of noninteracting $\pi$ electrons in graphene.
\cite{Wunsch_2006,Hwang_2007}
Moreover, in Eq.~(\ref{Eloss_ave}) we have introduced the fluctuation in an effective areal (or surface-projected) number density of external particles, $\delta\!\cN(\br)$, which is defined via its 2D FT as
$\DFcN(\bq)=\FcN(\bq)-\langle\FcN(\bq)\rangle$, with
\begin{eqnarray}
\FcN(\bq)\equiv\frac{1}{e}\int dz\,\psi(q,z)\,\Frho_0(\bq,z)=\sum_{j=1}^N\cF_j(\bq)\,\mathrm{e}^{-i\bq\cdot\br_j},
 \label{FcN}
\end{eqnarray}
where
\begin{eqnarray}
\psi(q,z)=\frac{\FGo(q;z_g,z)}{\FGo(q;z_g,z_g)}
 \label{psi}
\end{eqnarray}
is a profile function that takes into account the decay of the Coulomb interaction throughout the system with increasing distance from graphene, and
\begin{eqnarray}
\cF_j(\bq)=\int dz\,\psi(q,z)\,\FD_j(\bq,z-z_j)
 \label{Ff}
\end{eqnarray}
may be considered to be a weighted form factor of the $j$th particle.

\subsection{Friction regime and conductivity of graphene}

In order to use the ELM to obtain the DC conductivity of graphene, we require an ensemble average of the energy loss rate to the lowest order in speed $v$, which corresponds to the friction regime for slowly moving external charges. This is easily accomplished by expanding the loss function $\Im\left[-1/\ve(q,\omega)\right]$ in Eq.~(\ref{Eloss_ave}) to the leading order in frequency by using the truncated expansion for the polarization function of doped graphene,\cite{Allison_2010}
\begin{eqnarray}
\chi(q,\omega)= \chi_s(q)+\frac{i\omega}{\pi\hbar v_F^2}\,\cU\!\!\left(2k_F-q\right)\sqrt{\left(\frac{2k_F}{q}\right)^2-1},
\label{loss}
\end{eqnarray}
where $\chi_s(q)=\chi(q,0)$ is the static polarization function and $k_F=\sqrt{\pi\barn}$ is an average value of the Fermi wavenumber for Dirac electrons in graphene.
We further define the auto-correlation function of charged impurities in Eq.~(\ref{Eloss_ave}) by
\begin{eqnarray}
\cS(q)\equiv\frac{1}{N}\left\langle\DFcN(-\bq)\DFcN(\bq)\right\rangle,
\label{cS}
\end{eqnarray}
and note that it only depends on the magnitude $q=\|\bq\|$ when the distribution of impurities is isotropic in the directions parallel to graphene.
This allows us to finally obtain from Eq.~(\ref{Eloss_ave})
\begin{eqnarray}
\left\langle\frac{dW}{dt}\right\rangle=2\hbar k_F N v^2 r_s^2 \int\limits_0^{2k_F}
\frac{dq\,\sqrt{1-\left(q/2k_F\right)^2}}{\left[\ebg(q)+4k_Fr_s/q\right]^2}\,\cS(q),
\label{Eloss_final}
\end{eqnarray}
where $r_s=e^2/\left(\hbar v_F\right)\approx 2$ with $v_F$ being the Fermi speed of Dirac electrons. Note that the quantity $F_s\equiv\left\langle dW/dt\right\rangle/v$ is an average total stopping, or drag force that acts on the moving system of external charges,
\cite{Radovic_2012} which may be used to, e.g., evaluate the friction coefficient $\eta$ for an adsorbed layer on graphene from the expression $\eta=F_s/v$ in the limit $v\rightarrow 0$.\cite{Allison_2010,Krim_2012}

Within the ELM, by reversing the frames of reference one may express the energy loss rate in graphene by the standard expression of classical electrodynamics,
\begin{eqnarray}
\left\langle\frac{dW}{dt}\right\rangle=\int d^2\br\,\left\langle\bJ\cdot\bE\right\rangle,
\label{Eloss_graphene}
\end{eqnarray}
where $\bJ=\sigma\bE$ is the current density of  charge carriers in graphene, induced by a constant electric field  $\bE$ applied across graphene, and $\sigma$ is its DC conductivity. Assuming a uniform charge carrier density $\barn$ across graphene, we may write $\bJ=-e\barn\bv$ in a steady-state regime, which gives $\left\langle dW/dt\right\rangle=\cA\left(e\barn v\right)^2/\sigma$, where $\cA$ is the macroscopic area of graphene.
We discard possible contribution to the conductivity of graphene coming from charge carrier scattering on short-ranged impurities, and we limit our considerations to sufficiently low temperatures to be able to neglect the contribution from scattering on phonons.
Thus, the final expression for the DC conductivity takes a form that is familiar from the SBT for doped graphene,\cite{Sarma_2011,Li_2011}
\begin{eqnarray}
\sigma=\frac{e^2}{h}\frac{\frac{\barn}{\nimp}}{2\int\limits_0^{1}du\,\frac{u^2\sqrt{1-u^2}}{\left[2+\frac{u}{r_s}\ebg(2k_Fu)\right]^2}
\cS(2k_Fu)},
\label{sigma}
\end{eqnarray}
where $\nimp=N/\cA$ is the mean areal number density of external charged particles.

\subsection{Variance of the potential in graphene and minimum conductivity}

Equation (\ref{sigma}) implies that the conductivity obtained within the SBT theory as a function of the average equilibrium charge carrier density in graphene, $\sigma(\barn)$, should vanish in a linear manner close to the neutrality point, i.e., when $\barn\rightarrow 0$, as long as $\ebg(0)$ and $\cS(0)$ remain finite. However, experiments show that the conductivity reaches a minimum value $\smin$ at the neutrality point due to electron-hole puddles in the charge carrier density across graphene, which are caused by fluctuations of the electrostatic potential in the plane of graphene due to spatial inhomogeneity of the external charged impurities.
\cite{Chen_2008,Tan_2007}
An estimate of $\smin$ may be found according to the SCT theory as $\smin=\sigma(n^*)$, where $n^*$ is referred to as a residual charge carrier density that gives a measure of the width of the plateau near the neutrality point where the conductivity minimum is reached.\cite{PNAS_2007} It was shown that
$n^*$ may be found as a solution of an equation involving the square of graphene's Fermi energy, $\varepsilon_F=\hbar v_F k_F$, and the variance of the fluctuating electrostatic potential in graphene, $\Dphi_g(\br)\equiv\left.\DPhi(\br,z)\right|_{z=z_g}$, that arises from a distribution of immobile external charges,
\begin{eqnarray}
(\hbar v_F)^2\pi\barn=C_0(\barn),
\label{SC}
\end{eqnarray}
where $C_0\equiv e^2\left\langle\Dphi_g^2(\br)\right\rangle$. We note that the SCT theory extends the applicability of the SBT result for the conductivity of graphene $\sigma(\barn)$ down to lower charge carrier densities with typically $n^*\lesssim 10^{11}$ cm$^{-2}$.\cite{PNAS_2007}

Working in the time-independent regime, we use the 2D spatial FT to express the fluctuating potential in graphene in terms of the 2D FT of the fluctuating charge density $\DFrho(\bq,z)\equiv\DFrho_0(\bq,z)$ as
\begin{eqnarray}
\DFphi_g(\bq)&=&\int\limits_{-\infty}^\infty \frac{\FGo(q;z_g,z)}{1+e^2\chi_s(q)\FGo(q;z_g,z_g)}\,\DFrho_0(\bq,z)\,dz
\label{FphiG}\\
&=&\frac{2\pi e}{q}\frac{\DFcN(\bq)}{\ve_s(q)},
\label{Fphizero}
\end{eqnarray}
where $\ve_s(q)=\ebg(q)+V_C(q)\,\chi_s(q)$ is the total dielectric function of the entire system in the static limit.
By invoking the translational invariance of the distribution of external charges in the directions of $\br$, we may use a general relation,
\begin{eqnarray}
\left\langle\DFcN(\bq')\DFcN(\bq)\right\rangle
=\nimp\,\delta\!(\bq'+\bq)\,\cS(\bq),
\label{trans}
\end{eqnarray}
that allows us to write
\begin{eqnarray}
C_0=\nimp\int\frac{d^2\bq}{(2\pi)^2}\,\left[\frac{V_C(q)}{\ve_s(q)}\right]^2\cS(q).
\label{C0}
\end{eqnarray}

\subsection{Statistical description of external charges}

It is important to make distinction between the geometric structure of the external particle system and the statistical distribution of the charge density functions $\Delta_j(\bR)$ for individual particles.
Assuming that those two characteristics of the system are statistically independent,
the geometric structure may be modeled by using the one- and two--particle distribution functions for their positions
\begin{eqnarray}
F_1(\br,z)=\frac{N}{\cA}f_1(z),
\label{F_1}
\end{eqnarray}
and
\begin{eqnarray}
F_2(\br_1,\br_2;z_1,z_2)&=&\frac{N(N-1)}{\cA^2}f_1(z_1)f_1(z_2)
\nonumber\\
&&\times
g(\br_2-\br_1;z_1,z_2),
\label{F_2}
\end{eqnarray}
where $f_1(z)$ describes the distribution of particle positions along the $z$ axis and is normalized to one,
whereas $g(\br;z_1,z_2)$ is the usual pair correlation function.
A significant further simplification may be achieved by assuming that the charge densities of individual particles are identically distributed, so that $\Delta_j(\bR)=\Delta(\bR)$ for all $j=1,2,\ldots,N$. Still, Eqs.~(\ref{FcN}) and (\ref{Ff}) show that the corresponding individual particle form factors generally remain entangled with the $z$ dependence of the geometric arrangement of particle positions, unless all the particles reside in the same plane, say $z=z_0$.

Accordingly, we first consider a 2D geometric model with $f_1(z)=\delta(z-z_0)$, which is commonly used in all theoretical modelings of the effects of correlated charged impurities on the conductivity of graphene.\cite{Sarma_2011,PNAS_2007,Yan_2011,Li_2011} In that case, we find that the auto-correlation function from Eq.~(\ref{cS}) may be written as
\begin{eqnarray}
\cS(\bq)=
\left\langle\left|\cF_0(\bq)\right|^2\right\rangle-
\left|\left\langle\cF_0(\bq)\right\rangle\right|^2+
\left|\left\langle\cF_0(\bq)\right\rangle\right|^2S_{2D}(q),
\label{cS_2D}
\end{eqnarray}
where each particle is characterized by an "atomic" form factor
\begin{eqnarray}
\cF_0(\bq)=\int dz\,\psi(q,z)\,\FD(\bq,z-z_0),
 \label{Ff0}
\end{eqnarray}
and
\begin{eqnarray}
S_{2D}(\bq)= 1+\nimp\int d^2\br\,\mathrm{e}^{i\bq\cdot\br}\left[g_{2D}(\br)-1\right]
 \label{S2D}
\end{eqnarray}
is a "geometric" structure factor that describes the arrangement of external particles in the plane $z=z_0$. As regards the corresponding pair correlation (or radial distribution) function $g_{2D}(\br)=g_{2D}(r)$,
in addition to uncorrelated particles with $g_{2D}(r)=1$, we consider two models that contain a single parameter $r_c$ characterizing the inter-particle correlation distance: a step-correlation (SC) model with $g_{2D}(r)=\cU(r-r_c)$, where $\cU$ is a Heaviside unit step function, which was often used in the previous studies of charged impurities in graphene,\cite{Yan_2011,Li_2011} and the HD model, in which particles interact with each other as hard disks of the diameter $r_c$.
\cite{Rosenfeld_1990}

There are several advantages to using the HD model over the SC model.
First, the former model is based on a Hamiltonian equation for the thermodynamic state of a 2D fluid with a well-defined pair potential between impurities, whereas the latter model is an \emph{ad hoc} description of the impurity distribution, made-up for simple, analytic results. That is not to say that the SC model is poor at capturing the interesting results in the conductivity of graphene with correlated impurities.\cite{Yan_2011,Li_2011}
However, from Eq.~(\ref{sigma}) it is obvious that,  with $k_F=\sqrt{\pi\barn}$, the initial slope of $\sigma(\barn)$ is strongly influenced by the limiting value of the structure factor $\cS(q)$ as $q\rightarrow 0$, that is, by the value of $S_{2D}(0)$ via Eq.~(\ref{cS_2D}). It is well known that $S_{2D}(0)$ is related to the isothermal compressibility of a 2D fluid,\cite{Hansen_1986} which may be expressed as a function of the packing fraction defined by $p=\pi\nimp r_c^2/4$. Thus, $p$ is a key measure of performance of the two models.
It was recently shown by Li \emph{et al.}\cite{Li_2011} that the SC model gives reliable results for packing fractions $p\ll 1$ by comparing the analytical result for the 2D structure factor in that model, $\SSC(q)$, with a numerically calculated structure factor of a hexagonal lattice of impurities. However, the analytical limit $\SSC(0)=1-4p$ shows that the SC model already breaks down for $p\ge 0.25$ because the corresponding compressibility becomes negative at higher packing fractions.
On the other hand, it was recently shown that the interaction potential between two point ions near doped graphene is heavily screened and, moreover, exhibits Friedel oscillations with inter-particle distance, giving rise to a strongly repulsive core region of distances on the order of $k_F^{-1}$ that resembles the interaction among hard disks with diameter $r_c\sim k_F^{-1}$.\cite{Radovic_2012}
Therefore, we may estimate that the packing factor could reach values on the order $p\sim\nimp/\barn$ that may not always be negligibly small, necessitating the use of a model that goes well beyond the SC model, at least for systems of adsorbed alkali-atom submonolayers on graphene.\cite{Yan_2011}
In that respect, we note that various parameterizations of the HD model extend its applicability to include phase transitions in a 2D fluid as a function of the packing fraction, \cite{Mak_2006} even going up about $p=0.9$, corresponding to a crystalline closest packing where hard disks form a hexagonal structure in 2D.\cite{Guoa_2006}
In this work, we use a simple analytical parametrization for the 2D structure factor in the HD model,  $\SHD(q)$,  provided by Rosenfeld\cite{Rosenfeld_1990}
(see Appendix B) which works reasonably well for packing fractions up to about $p=0.69$, just near the freezing point of a 2D fluid.

Regarding the structure of individual charged particles within the 2D geometric model, we study a few specific examples.
First we consider a point particle of charge $Ze$ that carries a dipole moment $\bmu$ with the density function
\begin{eqnarray}
\Delta_{\mathrm{p}}(\bR)=\left(Z-\bD\!\cdot\!\nabla_{\bR}\right)\,\delta\!\left(\bR\right),
\label{point}
\end{eqnarray}
where $\bD=\bmu/e$ is an effective dipole length and $\delta\!(\bR)=\delta\!\left(\br\right)\delta\!\left(z\right)$ is a 3D delta function,
which gives a form factor from Eq.~(\ref{Ff0}) as
\begin{eqnarray}
\cF_{\mathrm{p}}(\bq)=\left(Z+i\,\bq\!\cdot\!\bD_\parallel\right)\psi(q,z_0)+D_\perp\left.\frac{\partial \psi(q,z)}{\partial z}\right|_{z=z_0},
 \label{Fpoint}
\end{eqnarray}
where $\bD_\parallel=\bmu_\parallel/e$ and $D_\perp=\mu_\perp/e$ are the effective dipole lengths in the directions parallel and perpendicular to graphene, respectively.
We note that, having in mind that the first two terms in the right-hand side of Eq.~(\ref{cS_2D}) represent the variance of the form factor $\cF_0(\bq)$, all of the three parameters of the point particle model, namely, $Z$, $\bD_\parallel$ and $D_\perp$ may exhibit fluctuations about their respective means (with the mean $\langle\bD_\parallel\rangle=0$ due to the presumed isotropy), as well as mutual cross-correlations. In addition, assuming $\nimp$ to be small enough, the perpendicular dipole moment component may depend on the local electrostatic field $E_\perp$ according to $\mu_\perp=\alpha E_\perp$, where $\alpha$ is an effective dipole polarizability near graphene.

We also consider a cluster of uniformly distributed charge $Ze$ within a disk of radius $\rcl$ parallel to graphene with
\begin{eqnarray}
\Delta_{\mathrm{cl}}(\bR)=\frac{Z}{\pi\rcl^2}\,\cU\!\left(\rcl-r\right)\,\delta\!\left(z\right),
\label{cluster}
\end{eqnarray}
giving
\begin{eqnarray}
\cF_{\mathrm{cl}}(\bq)=
\frac{2Z}{q\rcl}\,J_1\!\left(q\rcl\right)\,\psi(q,z_0),
\label{Fcluster}
\end{eqnarray}
where $J_1$ is a Bessel function of order one. We limit our considerations to cases with $k_F\rcl\ll 1$, validating the perturbative treatment of charge carrier scattering on such clusters,\cite{Katsnelson_2010} and we also assume $\pi\nimp\rcl^2\ll 1$ to avoid the interference in scattering patterns from neighboring clusters.

On the other hand, it is of interest to explore the effects a fully $z$-dependent geometric structure of particle positions in 3D, with arbitrary distribution function
$f_1(z)$ and the pair correlation function that depends on the $z$ coordinates, $g_{3D}(\br_2-\br_1;z_1,z_2)$. In this case, we only consider point charges with $Z=1$ and obtain the auto-correlation function from Eq.~(\ref{cS}) as
\begin{eqnarray}
\cS(\bq)&=&\int dz\,f_1(z)\,\psi^2(q,z)+\int dz\,f_1(z)\,\psi(q,z) 
\nonumber \\
&&\times
\int dz'\,f_1(z')\,\psi(q,z')\left[S_{3D}(\bq;z,z')-1\right] ,
 \label{cS3D}
\end{eqnarray}
where partial structure factor in the 3D case is defined by
\begin{eqnarray}
S_{3D}(\bq;z,z')= 1+\nimp\int d^2\br\,\mathrm{e}^{i\bq\cdot\br}\left[g_{3D}(\br;z,z')-1\right].
 \label{S3D}
\end{eqnarray}
Any realistic modeling of the 3D pair correlation function in the presence of charged graphene is beyond the scope of the present study, so we only consider uncorrelated point charges with
$g_{3D}(\br;z,z')=1$, and focus on the effect of their distribution over the depth $z$. In a first study of this type, we only consider the case $f_1(z)=1/L$
for a homogeneous distribution of point charges throughout a dielectric slab of finite thickness $L$.
In Appendix B we also provide a result for semi-infinite region ($L\rightarrow\infty$) based on a pair correlation function $g_{3D}(R)$ for a bulk one-component plasma (OCP) with the volume density of charged particles $N_\mathrm{imp}=N/\left(AL\right)$, which may be of interest in future work.

\end{section}

\begin{section}{Results and discussion}

In this section, we study several special configurations of graphene with the surrounding dielectrics by using the electrostatic GF, which is derived in Appendix A for a three-layer structure of Fig.~1, defined on the intervals $I_1=[-L,0]$, $I_2=[0,H]$ and $I_3=[H,\infty)$ along the $z$ axis that are characterized by the relative bulk dielectric constants $\ve_j$ with $j=1,2,3$, respectively.

In Fig.~2 we consider a two-layer structure that consists of a semi-infinite SiO$_2$ substrate ($L\rightarrow\infty$ with $\ve_1=3.9$) and a semi-infinite layer of air ($H\rightarrow\infty$ with $\ve_2=1$, or $H=0$ with $\ve_3=1$) with graphene placed right on their boundary at $z_g=0$.
We show the dependence of graphene's conductivity $\sigma$ on its average charge carrier density $\barn$ for
a planar distribution of charged impurities with fixed $Z=1$ and no dipole moment, having the areal number density $\nimp=10^{12}$ cm$^{-2}$, which are all placed a distance $d$ away from graphene. We show the results for several values of the correlation length $r_c$ among the impurities, which are obtained by using the SC and the HD models for their 2D structure factor, and note that the SC model only yields physical results for $r_c<5.6$ nm for the given value of $\nimp$. In addition to the case of point-like impurities being placed directly on graphene ($d=0$ and $\rcl=0$), we also show in Fig.~2 the effects of point-like impurities embedded at $d=0.3$ nm inside the SiO$_2$ substrate, as well as disk-like impurities with fixed radius $\rcl=2$ nm placed on graphene ($d=0$).
\begin{figure}
\centering
\includegraphics[width=0.45\textwidth]{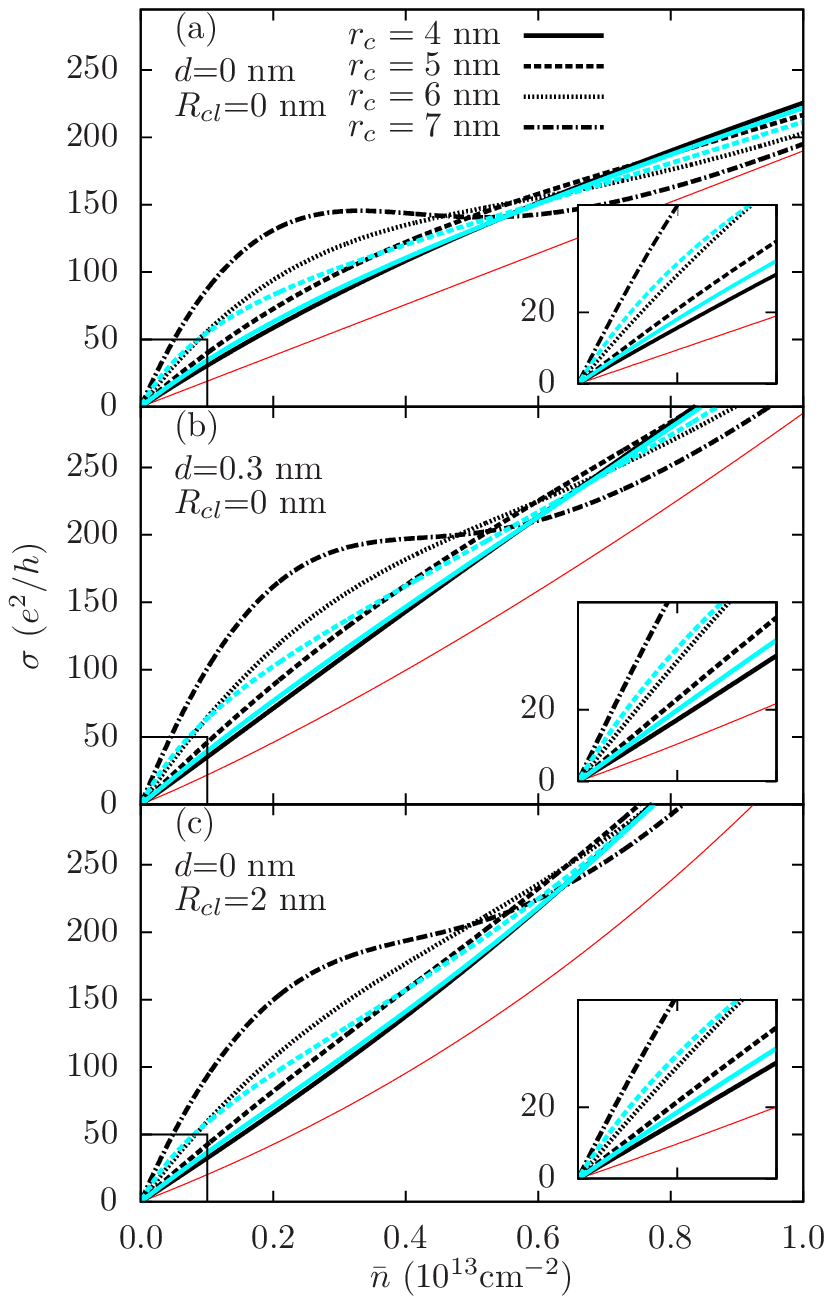}
\caption{(Color online)
The dependence of conductivity (in units of $e^2/h$) on the average charge carrier density $\barn$ (in units of $10^{13}$ cm$^{-2}$) for a two-layer structure that consists of a semi-infinite SiO$_2$ substrate ($L\rightarrow\infty$, $\ve_1=3.9$) and a semi-infinite layer of air ($H\rightarrow\infty$, $\ve_2=1$, or $H=0$, $\ve_3=1$), with zero gap between them and graphene placed on their boundary ($z_g=0$). A planar distribution of charged impurities with $Z=1$ and no dipole moment, having the areal number density $\nimp=10^{12}$ cm$^{-2}$ and the correlation distance $r_c$ between them, is placed a distance $d$ away from graphene. Results are shown for uncorrelated impurities [thin (red) solid lines], for the SC model with $r_c=$ 4 and 5 nm [thick solid and dashed gray (light blue) lines, respectively], and for the HD model with $r_c=$ 4, 5, 6, and 7 nm [thick black solid, dashed, dotted, and dash-dotted lines, respectively]. Panels (a) and (b) show the cases of point-like impurities on graphene ($d = 0$) and at $d=0.3$ nm in the SiO$_2$ substrate, respectively, whereas panel (c) shows disk-like impurities with the cluster radius $\rcl=2$ nm placed on graphene ($d=0$). The insets show the blow-ups of the regions with $\barn\leq 10^{12}$ cm$^{-2}$.}
\end{figure}

As regards the effects of finite $d$ and $\rcl$, one notices in Fig.~2 that they both contribute to an increase in the slope of conductivity at higher $\barn$ values, as expected, where they even give rise to a super-linear dependence of conductivity on $\barn$ for smaller values of the correlation length $r_c$. (Note that the case of uncorrelated disks with $r_c=0$ is somewhat unphysical as the disks are allowed to overlap.) However, the effects of finite $d$ and $\rcl$ are relatively weak and only affect quantitative details of conductivity at higher $\barn$, whereas comparison among the insets in Fig.~2 shows that their effects are barely noticeable at $\barn \lesssim 10^{12}$ cm$^{-2}$.

The most prominent effect in Fig.~2 is a strong increase of the initial slope of conductivity as a function of $\barn$ (and hence an increase in mobility of graphene, $\mu=\sigma/\left(e\barn\right)$) at low values of $\barn$ as the correlation length $r_c$ increases.
One notices from the insets in Fig.~2 that the initial slopes from the SC model are higher than those from the HD model for the same value of $r_c$ because $\SSC(0)<\SHD(0)$, but the latter model permits the use of much larger values of $r_c$ than the former model, hence giving rise to rather large initial slopes of the conductivity at the largest packing fractions shown. (Notice that the case with a maximum packing fraction of $p\approx 0.38$ that is shown in Fig.~2 is still well within the interval of confidence for the HD model used here.\cite{Rosenfeld_1990}) As the charge carrier density $\barn$ increases, the conductivity shows a sub-linear dependence on $\barn$ that becomes more pronounced as the correlation length $r_c$ increases. In the case of $d=0$ and $\rcl=0$ the sub-linear dependence occurs for all $r_c>0$, whereas in the cases of finite $d$ or $\rcl$ values the sub-linear dependence may even overcome the opposite effect of super-linear dependence for sufficiently large $r_c$s.
For the largest $r_c$ value shown in Fig.~2, the sub-linear behavior even gives rise to a pronounced saturation effect in the conductivity of graphene with increasing $\barn$, which is sometimes observed in experiments.\cite{Tan_2007,Yan_2011}
Thus, high packing fractions that result from long correlation distances among the charged impurities can give rise to both higher initial slope of conductivity at lower $\barn$ \emph{and} a more pronounced sub-linear dependence of conductivity at higher $\barn$ with the HD model than those that can be achieved with the SC model.
We pause to discuss those two effects in some detail.

Various models that attempt to reproduce the experimental dependence of graphene's conductivity $\sigma$ on its charge carrier density $\barn$
use the areal density of charged impurities $\nimp$ as free parameter to fit the slope of conductivity in the range of $\barn$ values where that dependence is found to be predominantly linear.
Ignoring the relatively narrow region of $\barn$ values around zero where the conductivity of a nominally neutral graphene reaches a minimum, one sees that Eq.~(\ref{sigma}) implies a linear dependence of conductivity in the form $\sigma=c\,\barn/\left[\nimp S_{2D}(0)\right]$ when $\barn\rightarrow 0$, where $c$ is constant when the dielectric media are semi-infinite. For a system of uncorrelated impurities that may be described as a 2D gas, one simply finds $\sigma=c\,\barn/\nimp$ because $S_{2D}(0)=1$. However, when impurities are strongly correlated, one should consider their number $N$ to be a random variable because different samples of graphene flakes with fixed area $A$ may cover different regions of a much larger area of the substrate plagued by varying concentrations of impurities. Then, the impurity density should be defined in terms of the average number of impurities covered by the graphene flake, $\nimp=\left\langle N\right\rangle/A$. On the other hand, the long wavelength limit of the structure factor may be expressed as the ratio $S_{2D}(0)=\left\langle \DN^2\right\rangle/\left\langle N\right\rangle$,
where the numerator is the variance in $N$,\cite{Hansen_1986} with $\DN=N-\left\langle N\right\rangle$ being the fluctuation in the number of impurities that are covered by the graphene flake.
Therefore, from the statistical point of view, the $\barn\rightarrow 0$ limit of the SBT conductivity should be reinterpreted as $\sigma=c\,\barn/\nimp^*$, where we define $\nimp^*=\left\langle \DN^2\right\rangle/A$ to be an effective density of impurities rather than the average density. In general, $\nimp^*\neq\nimp$ unless $N$ is Poisson distributed, i.e., the impurities behave as an ideal 2D gas. Clearly, the distinction between $\nimp^*$ and $ \nimp$ should be borne in mind when attempting to use $\nimp$ as a fitting parameter in modeling the slope of graphene's conductivity in the presence of a liquid-like distribution of charged impurities.

On the other hand, the sub-linear dependence of graphene's conductivity on $\barn$ at large doping densities is often modeled by combining
the scattering processes of its charge carriers on both charged impurities and short-ranged impurities via the Matthiessen's rule.\cite{Yan_2011}
However, the density of atom-size defects in graphene that could give rise to short-range scattering is extremely low due to the structural and compositional resilience of graphene's atomic lattice, so that ''the source of the proposed weak short-range scattering is mysterious.''\cite{Yan_2011}
Another contender for the explanation of the sub-linear conductivity is the resonant scattering model that invokes the existence of bound-state resonances in the $\pi$ electron bands due to chemisorbed molecules on graphene.\cite{Ferreira_2011} However, the fact that graphene is chemically inert also makes this mechanism unlikely in most situations.
On the other hand, it was recently shown that the charge carrier scattering on charged impurities in a substrate may also give rise to the sub-linear behavior of conductivity in highly doped graphene in the presence of a strong spatial correlation among the impurities.\cite{Yan_2011,Li_2011} Noting that the sub-linear behavior was demonstrated in simulations based on the SC model with small packing fractions,\cite{Yan_2011,Li_2011} we follow the same idea and suggest that, by being able to go to much larger packing fractions in the HD model than in the SC model, one may include large enough values of $r_c$ in simulations that could even give rise to saturation of graphene's conductivity at high enough charge carrier densities, thus eliminating the need to invoke the existence of resonance scatterers or atom-size defects in graphene.
Namely, one may verify that, with increasing packing fraction the structure factor $\SHD(q)$ develops a very pronounced peak at the wavenumber $q=q_\mathrm{shell}$ corresponding to the first coordination shell due to the nearest neighbors.\cite{Rosenfeld_1990,Guoa_2006} So, from Eq.~(\ref{sigma}) it follows that a relatively sudden increase in the value of the integral over $u$ may be expected with the HD model when $k_F$ surpasses the value $q_\mathrm{shell}/2\sim\pi/r_c$, causing a slowdown in the increase of $\sigma$ when $\barn\sim\pi/r_c^2$ that is reminiscent of the saturation in conductivity. For example, in the case of the largest correlation distance shown in Fig.~2, $r_c=7$ nm, one finds that a strong saturation of the conductivity indeed occurs at about $\nimp=\pi/r_c^2\approx 6.4\times 10^{12}$ cm$^{-2}$.

\begin{figure}
\centering
\includegraphics[width=0.5\textwidth]{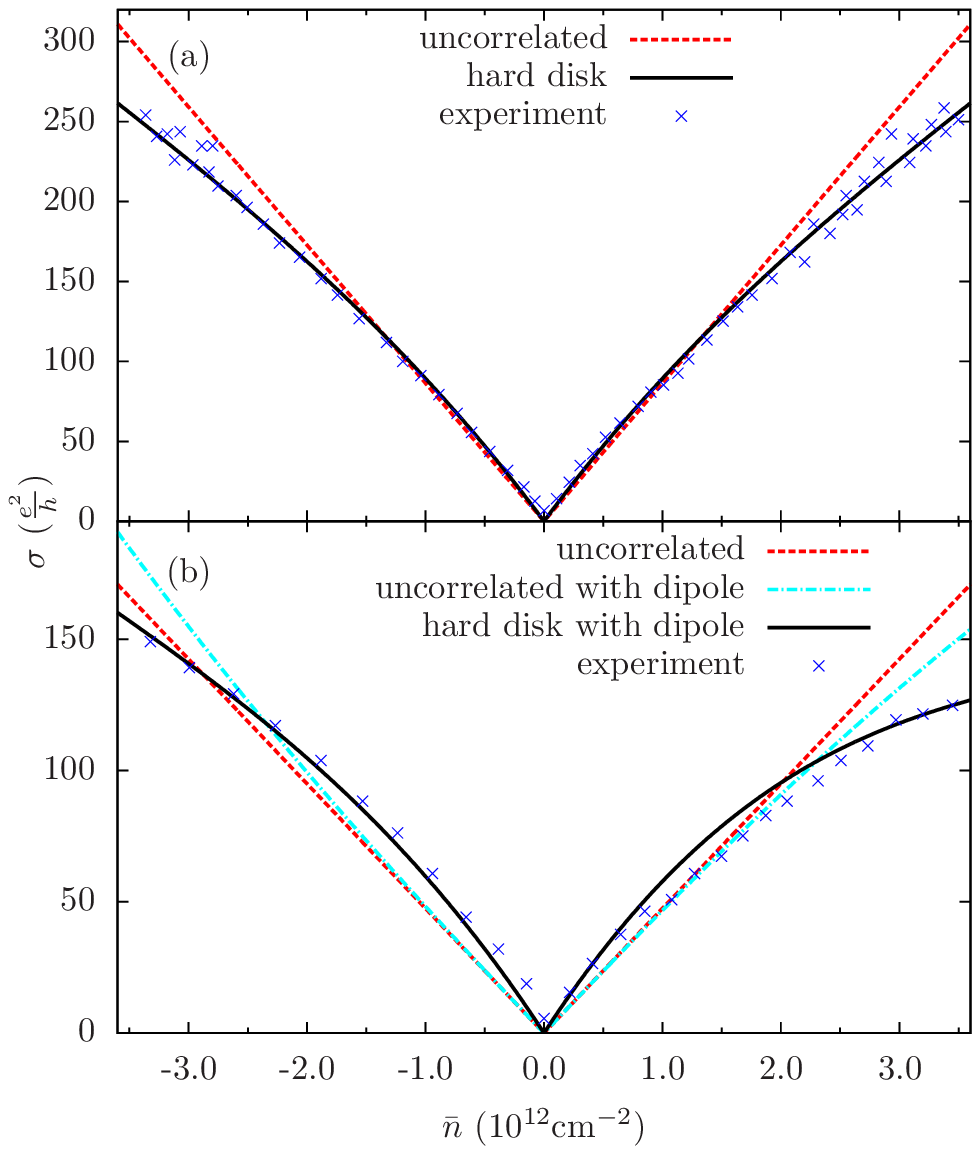}
\caption{(Color online)
The dependence of conductivity (in units of $e^2/h$) on the average charge carrier density $\barn$ (in units of $10^{12}$ cm$^{-2}$) for a two-layer structure that consists of a semi-infinite SiO$_2$ substrate ($L\rightarrow\infty$, $\ve_1=3.9$) and a semi-infinite layer of air ($H\rightarrow\infty$, $\ve_2=1$, or $H=0$, $\ve_3=1$), with zero gap between them and graphene placed on their boundary ($z_g=0$). A planar distribution of unit ($Z=1$) point-like charged impurities,
having the areal number density $\nimp$ and the correlation distance $r_c$ between them, is placed on graphene and
is allowed to have a non-zero perpendicular dipole moment with polarizability $\alpha$ per impurity.
The results from the HD model
(black solid lines) are fitted to the experimental data from Ref.~$\cite{Tan_2007}$ (symbols), with the best fit in panel (a) obtained for
$\nimp=3\times 10^{11}$ cm$^{-2}$ with $r_c=6.8$ nm (packing fraction $p=0.11$) and $\alpha=0$, and the best fit in panel (b) obtained for
$\nimp=7.4\times 10^{11}$ cm$^{-2}$ with $r_c=6.3$ nm ($p=0.23$) and $\alpha=1150$ \AA$^3$. Also shown are the results for uncorrelated impurities ($r_c=0$) with $\alpha=0$ on both panels [dashed gray (red) lines], as well as for the uncorrelated impurities ($r_c=0$) with $\alpha=1150$ \AA$^3$ in panel (b) [dash-dotted grey (light blue) line].
 }
\end{figure}
In Fig.~3 we consider the same configuration of single-layer graphene atop a semi-infinite SiO$_2$ substrate with a semi-infinite layer of air above it as in Fig.~2, and attempt to model the experimental data for conductivity versus charge carrier density $\barn$ from Ref.~$\cite{Tan_2007}$ by using the HD model
for a 2D distribution of point charges with $Z=1$.
We select two graphene samples from Ref.~$\cite{Tan_2007}$ labeled K17 and K12, which both exhibit sub-linear behavior with increasing $\barn$, with K17 being symmetric and K12 showing an electron-hole asymmetry (i.e., asymmetry with respect to the sign of $\barn$).
The physical mechanism(s) that occasionally give rise to this kind of asymmetry in graphene are still unclear, so we explore here the possibility that the presence of the perpendicular component of dipole moment in each impurity, $D_\perp$, may give rise to a sizeable asymmetry, as that seen in Fig.~3 for the sample K12. We assume $D_\perp=\alpha E_\perp/e$, where $\alpha$ is the effective polarizability and $E_\perp$ is the total perpendicular electric field near graphene. Assuming $\nimp$ to be small enough, we may neglect mutual depolarization among the impurities and simply write $E_\perp=4\pi e\barn/\ve_1$, with $E_\perp$ being positive (negative) for electron (hole) doping of graphene.\cite{Maschhoff_1994}
The two samples were fitted in Ref.~$\cite{Tan_2007}$ by assuming that the impurities reside in graphene ($d=0$) and are uncorrelated, and the optimal linear symmetric fits were found with $\nimp=2.2\times 10^{11}$ cm$^{-2}$ for K17 and with $\nimp=4\times 10^{11}$ cm$^{-2}$ for K12.
We also assume the impurities to lie in graphene ($d=0$), and we use $\nimp$, $r_c$ and $\alpha$ as fitting parameters. In the case of the symmetric K17, the best fit is found for $\nimp=3\times 10^{11}$ cm$^{-2}$ with $r_c=6.8$ nm ($p=0.11$) and $\alpha=0$, whereas for the asymmetric case of K12 the best fit is found for  $\nimp=7.4\times 10^{11}$ cm$^{-2}$ with $r_c=6.3$ nm ($p=0.23$) and $\alpha=1150$ \AA$^3$. Both fits obtained with the HD model in Fig.~3 are quite satisfactory as far as the sub-linear behavior of conductivity is concerned, and the relatively large values of packing fractions used in both cases suggest the necessity of using the HD rather than the SC model. On the other hand, a good fit in the asymmetric case can only be achieved with a rather large value of $\alpha$, which indicates that the dipole mechanism may not be the primary cause of the electron-hole asymmetry in conductivity, at least for the experimental setting of Ref.~$\cite{Tan_2007}$ However, we note that the effective polarizability $\alpha$ of a single impurity may be significantly increased by the presence of a nearby conducting surface.\cite{Maschhoff_1994}

In Fig.~4 we consider a structure that consists of a dielectric material of finite thickness $L$ (we choose HfO$_2$ with $\ve_1=22$) and a semi-infinite layer of SiO$_2$  (either $H\rightarrow\infty$ with $\ve_2=3.9$, or $H=0$ with $\ve_3=3.9$) with graphene placed right on their boundary at $z_g=0$. This configuration may represent the physical situation where single-layer graphene sits on a thick SiO$_2$ substrate (with typically $H\sim 300$ nm) and is top-gated through a thin layer of HfO$_2$ (with $L\lesssim 10$ nm).
We show the dependence of the conductivity $\sigma$ on charge carrier density $\barn$ for several model distributions of point charge impurities in the HfO$_2$ layer with fixed $Z=1$ and no dipole moment, having the areal number density $\nimp=10^{12}$ cm$^{-2}$. We consider a homogeneous 3D distribution of uncorrelated charges throughout the HfO$_2$, which extends up to a distance $d$ from graphene, as well as a 2D planar distribution placed in HfO$_2$ a distance $d$ away from graphene, with both uncorrelated ($r_c=0$) and correlated ($r_c=6$ nm, $p\approx 0.28$) charges that are described with the HD model.

One notices in Fig.~4 that finite thickness $L$ exhibits strong effects on conductivity, both in quantitative and qualitative aspects, which are dependent on the underlying structure of charged impurities.
First noted is that the overall conductivity is generally increased compared to that seen in Figs.~1 and 2, which is expected due to the more efficient screening of charged impurities by a high-$\kappa$ material such as HfO$_2$. Moreover, the conductivity is seen to increase with decreasing $L$ for all $\barn$ in the 2D cases and only for lower $\barn$ in the 3D case, which may be explained by the more efficient screening of impurities due to the proximity of a metal gate.
Furthermore, the conductivity is larger in the 3D case than in the corresponding uncorrelated 2D case because the same number of impurities is spread over larger distances from graphene so that the resulting scattering potential in graphene is weaker.
As regards the distance $d$, one notices similar trends as in Fig.~2, namely, a finite $d$ increases both the value of conductivity and its slope (i.e., mobility) in both 3D and 2D models. However, as regards the effects of finite correlation length $r_c$ in the 2D models with finite $L$,
one sees little evidence to the increase in the initial slope of conductivity at lower $\barn$, in contrast to the trends seen in Fig.~2, whereas saturation of conductivity at higher $\barn$ seems to get stronger than in Fig.~2 as $L$ decreases. In fact, for the shortest thickness of $L=1$ nm for both $d=0$ and $d=0.3$ nm, this saturation turns into a broad maximum of conductivity around $\barn=10^{11}$ cm$^{-2}$, followed by a still broader minimum at higher $\barn$ values.

\begin{figure*}[h,t,p]
\centering
\includegraphics[width=0.9\textwidth]{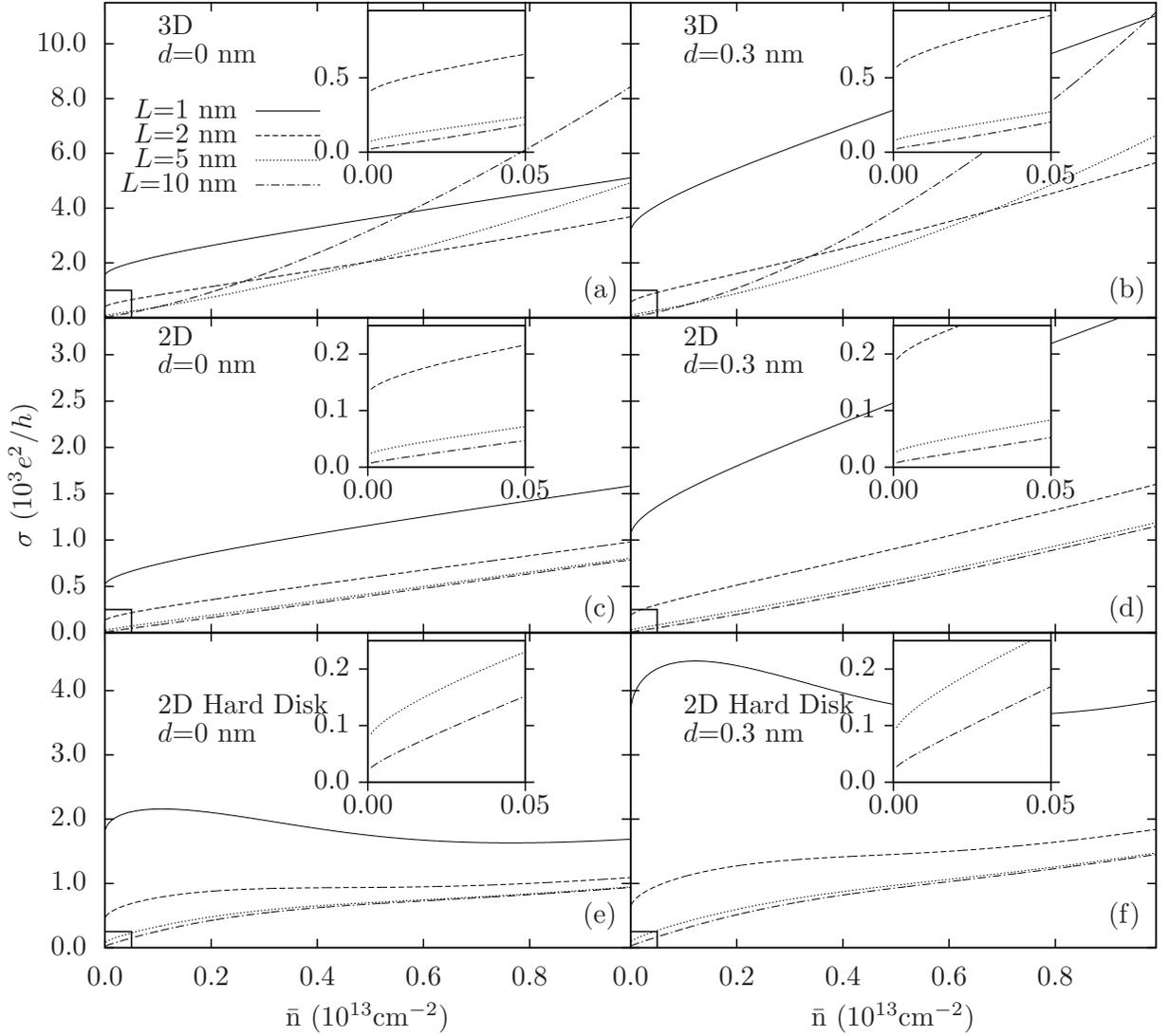}
\caption{
The dependence of conductivity (in units of $e^2/h$) on the average charge carrier density $\barn$ (in units of $10^{13}$ cm$^{-2}$) for a two-layer structure that consists of a HfO$_2$ ($\ve_1=22$) with finite thickness $L$ and a semi-infinite layer of SiO$_2$ ($H\rightarrow\infty$, $\ve_2=3.9$, or $H=0$, $\ve_3=3.9$)
with zero gap between them and graphene placed on their boundary ($z_g=0$).
The structure of the system of unit ($Z=1$) point-like charged impurities with no dipole moment, having the areal number density $\nimp=10^{12}$ cm$^{-2}$, is assumed to be either (a,b) a 3D homogeneous distribution throughout the HfO$_2$ layer extending up to a distance $d$ from graphene, or a planar 2D distribution placed in the HfO$_2$ layer a distance $d$ away from graphene, with the correlation distance being (c,d) $r_c=0$ or (e,f) $r_c$= 6 nm (giving the packing fraction $p\approx 0.28$ within the HD model). In panels (a,c,e) we set $d=0$, while in panels (b,d,f) we set $d=0.3$ nm.  The thickness of the HfO$_2$ layer takes values $L$ = 1 nm (solid lines), 2 nm (dashed lines), 5 nm (dotted lines), and 10 nm (dash-dotted lines). The insets show the blow-ups of the regions with $\barn\leq 5\times 10^{11}$ cm$^{-2}$.
 }
\end{figure*}
One remarkable feature seen in Fig.~4 is that the conductivity generally does not vanish in the SBT limit when $\barn\rightarrow 0$ for finite $L$, but rather reaches a minimum value $\sigma(0)$. This minimum may be easily estimated for $d=0$ by using the limiting form of the background dielectric constant $\ebg(q)=\ve_1/\left(2qL\right)$ when $qL\ll 1$ in Eq.~(\ref{sigma}), which then gives
\begin{eqnarray}
\sigma(0)=\left(\frac{\ve_1}{\pi r_sL}\right)^2\frac{e^2/(2h)}{\nimp\cS(0)}=\frac{4v_F}{\pi r_s}\frac{C_L^2}{\nimp\cS(0)},
\label{sigma_0}
\end{eqnarray}
where $\cS(0)=1/3$ for the 3D case, $\cS(0)=1$ for the uncorrelated 2D case, and $\cS(0)=\SHD(0)=(1-p)^3/(1+p)\approx 0.29$ for the correlated 2D case in the HD model.
In the second expression for $\sigma(0)$ in Eq.~(\ref{sigma_0}) we emphasize that the minimum conductivity in the SBT limit for neutral graphene is governed by the geometric capacitance per unit area, $C_L=\ve_1/(4\pi L)$, of the dielectric with finite thickness $L$ used in top-gating the graphene.

Finally, one notices in Fig.~4 that, as the thickness $L$ increases in the 3D case, the conductivity gains quite strong super-linear dependence with increasing $\barn$. This dependence may be estimated
by considering Eq.~(\ref{sigma}) in the limit of large but finite $L$, such that $qL\gg 1$. In that case, the background dielectric constant becomes $\ebg\approx\left(\ve_1+\ve_2\right)/2$, whereas the 3D structure factor, which is determined by the first term in Eq.~(\ref{cS3D}), goes as $\cS(q)\approx 1/\left(2qL\right)$, so that Eq.~(\ref{sigma}) gives $\sigma\propto\barn^{3/2}/N_\mathrm{imp}$, where $N_\mathrm{imp}=N/(AL)$ is the volume density of charge impurities.
We note that this behavior of conductivity in graphene at large $\barn$ is a consequence of the 3D nature of a distribution of uncorrelated charges that gives rise to the special form of structure factor, $\cS(q)\approx 1/\left(2qL\right)$. The lack of experimental observations of such super-linear dependence of conductivity in graphene should not be taken as evidence to rule out the role of 3D distributions of impurities, because both the correlation among impurities, as that described in the Appendix B for a OCP, as well as their clustering close to graphene seem to be capable of eliminating the super-linear dependence.

\begin{figure}
\centering
\includegraphics[width=0.48\textwidth]{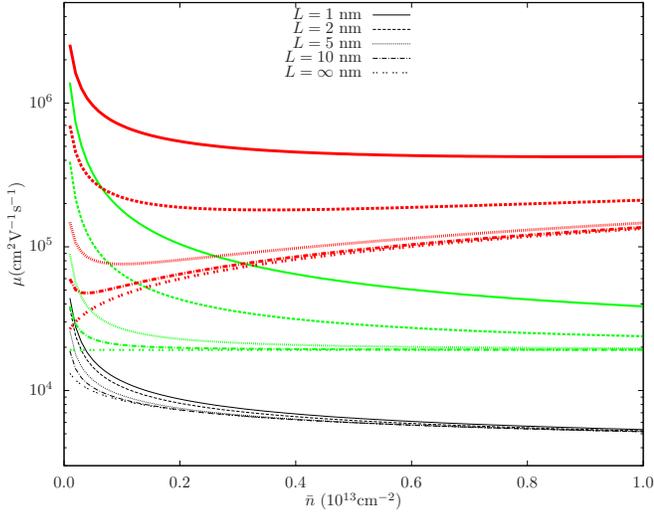}
\caption{(Color online)
The dependence of the mobility $\mu=\sigma/\left(e\barn\right)$, (in units of cm$^2$V$^{-1}$s$^{-1}$) on the average charge carrier density $\barn$ (in units of $10^{13}$ cm$^{-2}$) for a three-layer structure that consists of a HfO$_2$ ($\ve_1=22$) with thickness $L$, a layer of air ($\ve_2=1$) with thickness $H$, and a semi-infinite layer of SiO$_2$ ($\ve_3=3.9$), with graphene placed at distance $z_g$ above the top surface of the HfO$_2$ layer. A planar distribution of uncorrelated unit ($Z=1$) point-like charged impurities with no dipole moment, having the areal density $\nimp=10^{12}$ cm$^{-2}$, is placed a distance $d$ underneath graphene.
The cases of graphene with equal air gaps of $z_g=H-z_g=$ 0.3 nm towards the two dielectrics are shown with the impurities placed on graphene ($d=0$) [thick red (dark gray) lines] or on the top surface of the HfO$_2$ layer ($d=0.3$ nm) (thin black lines). The case of graphene with zero gaps ($z_g=H=0$) towards the two dielectrics and the impurities placed on graphene ($d=0$) [medium green (gray) lines] corresponds to the conductivity $\sigma$ shown Fig.~4(c). The thickness of the HfO$_2$ layer takes values $L$ = 1 nm (solid lines), 2 nm (dashed lines), 5 nm (dotted lines), 10 nm (dash-dotted lines), and $\infty$ (double-dotted lines).
 }
\end{figure}
In Fig.~5 we consider a three-layer structure that consists of a HfO$_2$ layer ($\ve_1=22$) with finite thickness $L$, a layer of air ($\ve_2=1$) of thickness $H=0.6$ nm, and a semi-infinite layer of SiO$_2$ ($\ve_3=3.9$), with graphene placed in the air at $z_g=0.3$ nm, midway between the two dielectrics. This configuration is similar to that in Fig.~4 with graphene sandwiched between the HfO$_2$ and SiO$_2$ dielectrics, but we introduce in Fig.~5 gaps of air of equal thickness 0.3 nm on both sides of graphene. We investigate the effects of finite thickness $L$ on the mobility of graphene, $\mu=\sigma/\left(e\barn\right)$, as a function of charge carrier density $\barn$ for a 2D planar distribution of uncorrelated point charges with $Z=1$ and no dipole moment, having the areal density $\nimp=10^{12}$ cm$^{-2}$. We consider three configurations, with the impurities placed either (A) on graphene ($d=0$) or (B) on the surface of the HfO$_2$ layer a distance $d=0.3$ nm away from graphene, both in the presence of the 0.3 nm gaps, as well as the case (C) from Fig.~4(c) having zero gaps between graphene and the HfO$_2$ and SiO$_2$ dielectrics with the 2D distribution of uncorrelated charges placed on graphene ($d=0$). One may see in Fig.~5 that the mobility generally increases with decreasing $L$ within each of the three configurations, (A), (B) and (C), but that there are remarkable differences between them in the magnitude of the mobility and its dependence on $\barn$. In the configurations (A) and (C) with charge impurities placed on graphene, the mobility generally decreases with increasing $\barn$, whereas in the configuration (B) with the impurities placed on the surface of the HfO$_2$ layer with a finite gap relative to the graphene, the mobilities with higher $L$ values pass through a minimum at a low $\barn$ value and further increase as $\barn$ increases.
Moreover, the magnitudes of the mobility with equal $L$ values are seen in Fig.~5 to increase in the order of configurations (A)$\rightarrow$(C)$\rightarrow$(B), which is also the order of increasing spread of the curves with different $L$ values within each configuration. Finally, it is interesting to notice that differences between the magnitudes of the mobility in the three different configurations with $L\rightarrow\infty$ become diminished as $\barn$ decreases.

One may conclude from Fig.~5 that the existence of a finite gap between graphene and the nearby dielectric, as well as the precise location of impurities
within that gap (with the extreme positions being on graphene and on the surface of the dielectric) both have decisive influences on the mobility. Noting that the configuration (A) with impurities on graphene in the presence of finite gaps was considered in Ref.\cite{Ong_2012}, it is remarkable how closing the gaps increases the magnitude of the mobility and increases the spread of its values for different $L$ values, whereas moving the impurities to the surface of a HfO$_2$ layer in the presence of finite gaps further accentuates those two effects, and even gives rise to a non-monotonous dependence of the mobility on $\barn$ for thicker HfO$_2$ layers. While the role of the distance of impurities from graphene  was discussed in detail for the case of zero gaps,\cite{PNAS_2007} one may conclude from our analysis that the size of the gap(s) between graphene and the nearby dielectric(s) plays equally important role in modeling the conductivity of graphene in a broad range of charge carrier densities.

We next turn to studying the conductivity minimum as $\barn\rightarrow 0$ due to the presence of electron-hole puddles by using Eqs.~(\ref{SC}) and (\ref{C0}) based on the SCT theory.\cite{PNAS_2007} We only consider a 2D planar distribution of point charges with $Z=1$ having no dipole moment and note that, unlike the integral in Eq.~(\ref{sigma}) for conductivity, in order to render the integral in Eq.~(\ref{C0}) convergent one must assume that charged impurities are placed a finite distance $d$ away from graphene.

\begin{figure*}
\centering
\includegraphics[width=0.8\textwidth]{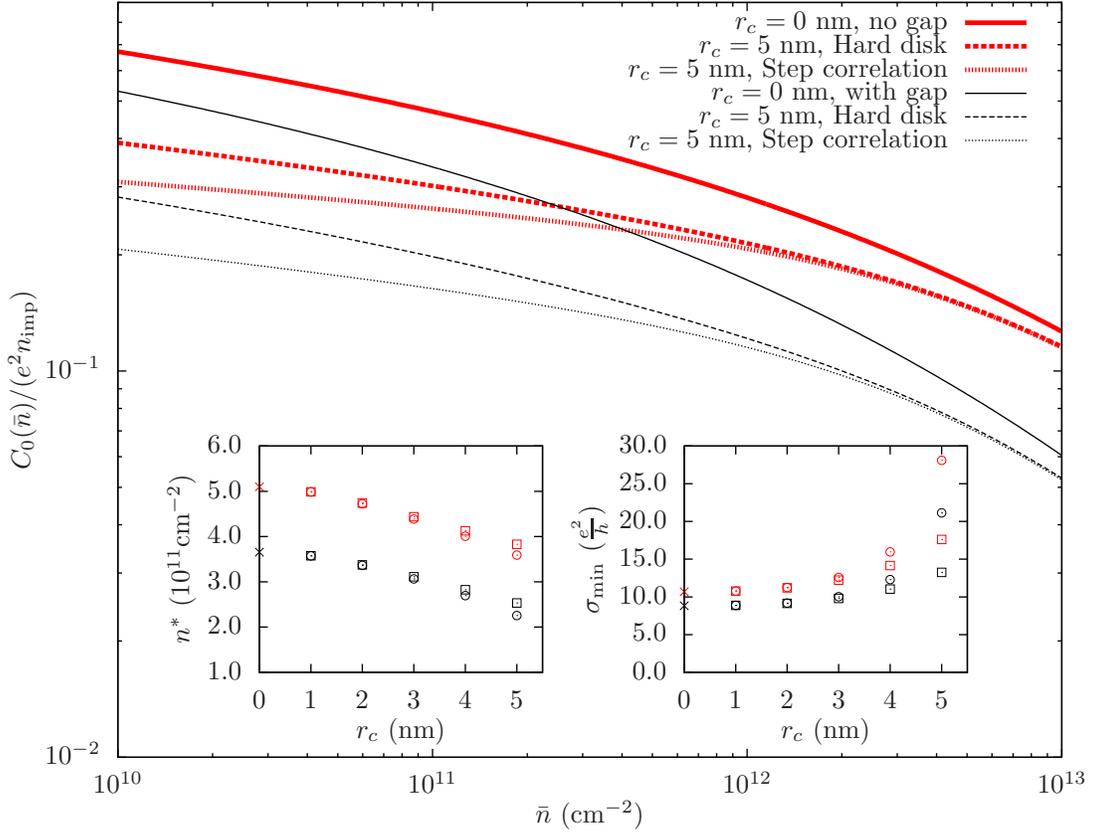}
\caption{(Color online)
The dependence of the variance of the potential in graphene $C_0$ (in units $e^2\nimp$) on the average charge carrier density $\barn$ (in units of cm$^{-2}$) for a two-layer structure that consists of a semi-infinite SiO$_2$ substrate ($L\rightarrow\infty$, $\ve_1=3.9$) and a semi-infinite layer of air ($H\rightarrow\infty$, $\ve_2=1$), with
graphene placed either on SiO$_2$ with zero gap ($z_g=0$) [thick red (grey) lines and symbols] or above SiO$_2$ with the air gap of $z_g=$ 0.3 nm (thin black lines and symbols).
A planar distribution of unit ($Z=1$) point-like charged impurities with no dipole moment, having the areal number density $\nimp=10^{12}$ cm$^{-2}$ and the correlation distance $r_c$ between them, is placed in/on SiO$_2$ at a fixed distance $d=0.3$ nm below graphene.
The case of uncorrelated impurities ($r_c=0$) (solid lines, crosses) is compared in the main panel with the cases of correlated impurities with $r_c=$ 5 nm (packing fraction $p=0.2$) in the HD model (dashed lines, circles) and in the SC model (dotted lines, squares). The left inset shows the residual charge carrier density (in units of $10^{11}$ cm$^{-2}$) and the right inset shows the conductivity minimum $\sigma_\mathrm{min}$ (in units of $e^2/h$), as functions of the correlation distance $r_c$ (in nm).
 }
\end{figure*}
In Fig.~6 we consider a configuration similar to that in Fig.~2, with a semi-infinite SiO$_2$ substrate ($L\rightarrow\infty$ with $\ve_1=3.9$) and a semi-infinite layer of air ($H\rightarrow\infty$ with $\ve_2=1$), with graphene placed in the air at a distance $z_g\ge 0$ above SiO$_2$. We show in the main panel of Fig.~6 the $\barn$ dependence of the variance of the potential in the plane of graphene $C_0$ from Eq.~(\ref{C0}) for a 2D distribution of charged impurities with density $\nimp=10^{12}$ cm$^{-2}$ that are placed in/on SiO$_2$ at a fixed distance $d=0.3$ nm below graphene. Specifically, we explore the effects of the size of the gap between graphene and the SiO$_2$ substrate by considering both the zero gap case with $z_g=0$ (impurities embedded at the depth of 0.3 nm inside SiO$_2$) and the finite gap case with $z_g=0.3$ nm (impurities placed on the surface of SiO$_2$). In addition to considering uncorrelated impurities, we use a finite correlation length of $r_c=5$ nm ($p\approx 0.2$) allowing us to compare in the main panel the effects of the SC and the HD models on $C_0$.
In the insets of Fig.~6, we show the dependence of the residual charge carrier density $n^*$ and the corresponding minimum conductivity $\smin=\sigma(n^*)$ on $r_c$
for both the HD and the CS models, in the presence of both zero and finite gaps.

\begin{figure*}
\centering
\includegraphics[width=0.8\textwidth]{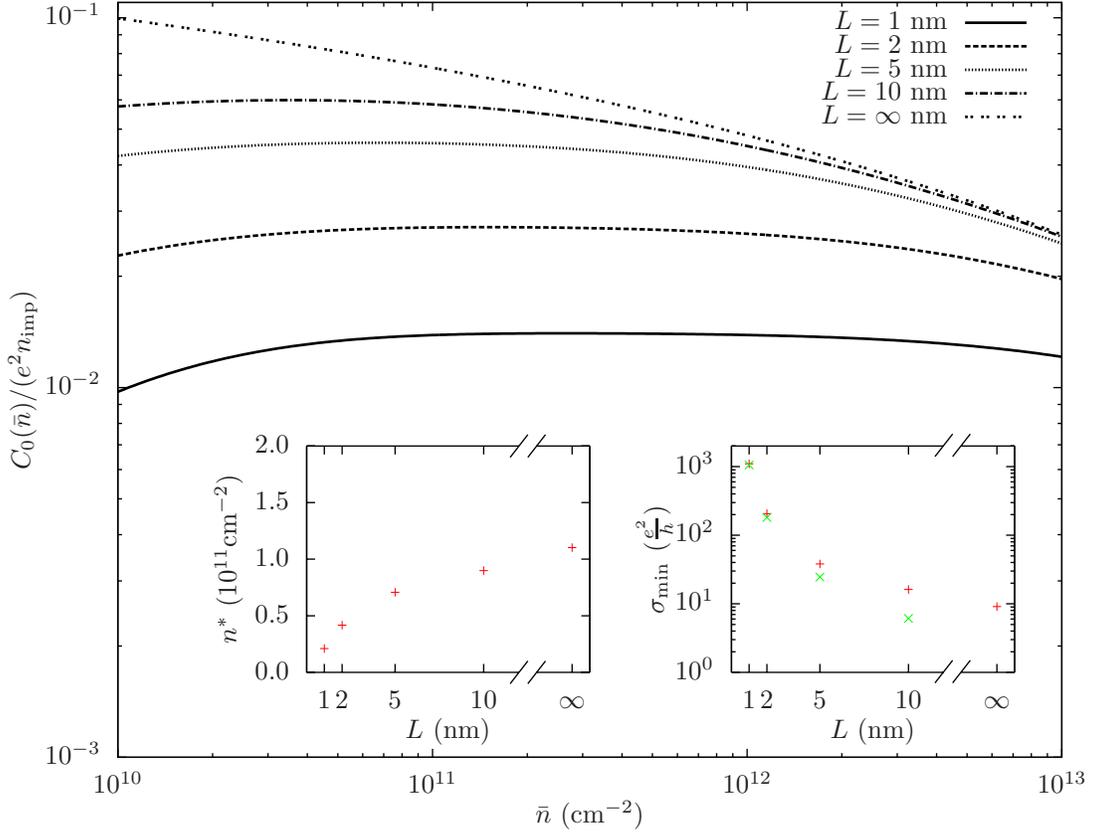}
\caption{(Color online)
The dependence of the variance of the potential in graphene $C_0$ (in units $e^2\nimp$) on the average charge carrier density $\barn$ (in units of cm$^{-2}$) for a two-layer structure that consists of a HfO$_2$ ($\ve_1=22$) with thickness $L$ and a semi-infinite layer of SiO$_2$ ($H\rightarrow\infty$, $\ve_2=3.9$, or $H=0$, $\ve_3=3.9$)
with zero gap between them and graphene placed on their boundary ($z_g=0$).
A planar distribution of uncorrelated unit ($Z=1$) point-like charged impurities with no dipole moment, having the areal number density $\nimp=10^{12}$ cm$^{-2}$ is embedded at a depth $d=0.3$ nm inside the HfO$_2$ layer, as in Fig.~4(d).
The thickness of the HfO$_2$ layer takes values $L$ = 1 nm (solid lines), 2 nm (dashed lines), 5 nm (dotted lines), 10 nm (dash-dotted lines), and $\infty$ (double-dotted lines). The (red) symbols $+$ show in the left inset the residual charge carrier density (in units of $10^{11}$ cm$^{-2}$) and in the right inset the conductivity minimum $\sigma_\mathrm{min}$ (in units of $e^2/h$), as functions of the thickness of the HfO$_2$ layer $L$. The (green) symbols $\times$ in the right inset show $\sigma(\barn=0)$ as a function of the HfO$_2$ layer thickness $L$.
}
\end{figure*}
One notices in Fig.~6 that the size of the gap between graphene and the SiO$_2$ substrate exerts a very strong effect on the magnitude of $C_0$ for all $\barn$, echoing similar conclusion drawn from the results analyzed in Fig.~5. The gap size also strongly affects the values of $n^*$ for all correlation lengths $r_c$, whereas the effect of the gap size on $\smin$ is seen to diminish as $r_c$ decreases. The latter result seems to justify the neglect of graphene--substrate gap, which is implicitly invoked in all simulations of the conductivity minimum in graphene in the presence of charged impurities with small or vanishing packing fractions.\cite{PNAS_2007,Yan_2011,Li_2011,Sarma_2011}
As far as the comparison between the HD and SC models is concerned, one sees a noticeable difference in the variance $C_0$ at small $\barn$, which diminishes at large $\barn$ values. The differences between the two models are surprisingly small in both $n^*$ and $\smin$, and only become noticeable when the packing fraction $p$ approaches the breakdown value of 0.25 for the SC model for sufficiently large correlation lengths $r_c$. These results again lend confidence to simulations that use the SC model with short correlation lengths among the charged impurities, which were seen to yield robustly satisfactory interpretations for the conductivity minimum in graphene due to electron-hole puddles.\cite{PNAS_2007,Yan_2011,Li_2011,Sarma_2011}

Finally, in Fig.~7 we consider a configuration that was studied in Fig.~4(d) with graphene sandwiched between a layer of HfO$_2$ of finite thickness $L$ and a semi-infinite layer of SiO$_2$, with no gaps between graphene and the two dielectrics, and with a 2D distribution of uncorrelated charged impurities of density $\nimp=10^{12}$ cm$^{-2}$ embedded at a depth $d=0.3$ nm inside the HfO$_2$ layer. In the main panel of Fig.~7 we show the dependence of the variance $C_0$ on the charge carrier density in graphene $\barn$, which exhibits an overall reduction in the magnitude of $C_0$ in comparison to Fig.~6 due to a larger dielectric constant of HfO$_2$, as well as a strong decrease of $C_0$ with decreasing $L$ owing to the screening of impurities by the nearby metallic gate. As a consequence, the resulting residual density $n^*$ is seen in an inset to Fog.~6 to decrease with decreasing $L$, which indicates that fluctuations in the charge carrier density in graphene due to electron-hole puddles would be gradually erased as the metal gate gets closer to graphene and provides more efficient screening of the fluctuations of the electrostatic potential. Finally, in the inset showing $\smin$ we explore the contribution of electron-hole puddles to raising the conductivity minimum above the SBT value $\sigma(0)$ that was discussed in Fig.~4 via Eq.~(\ref{sigma}) in the limit $\barn\rightarrow 0$.
It is interesting to note that, even though the contribution $\sigma(n^*)-\sigma(0)$  that comes from the residual density $n^*$ decreases with decreasing $L$, the dependence of $\sigma(0)\propto L^{-2}$ implied from Eq.~(\ref{sigma_0}) due to geometric capacitance of the HfO$_2$ layer appears to increase much faster with decreasing $L$, so that the net value of the conductivity minimum $\smin=\sigma(n^*)$ actually increases as the thickness $L$ of the HfO$_2$ layer decreases.

\end{section}

\begin{section}{Concluding remarks}

We have investigated the conductivity of doped single-layer graphene in the limit of semiclassical Boltzmann transport, as well as the conductivity minimum of a nominally neutral graphene within the Self-consistent transport (SCT) theory, placing emphasis on the effects due to the structure of charged impurities near graphene and the structure of the surrounding dielectrics. This was achieved by treating graphene as a zero-thickness layer embedded in a stratified structure of three dielectric layers and by using the full electrostatic Green's function for that structure.
We have used the Energy loss method to derive the conductivity of graphene from the friction force on a slowly moving structure of charged impurities, based on the polarization function of graphene within the RPA for its $\pi$ electrons treated as Dirac's fermions. Regarding the structure of charged impurities, we have analyzed the effects of their distance from graphene, the effects of correlation distance between the impurities within the hard-disk (HD) model for a 2D planar structure, and the effects of a homogeneous distribution of impurities over a 3D region. Besides point-charge impurities, we have analyzed the effects of a finite dipole moment on each impurity, as well as the effects of clustering of impurities into circular disks. Regarding the structure of the surrounding dielectrics, we have analyzed the effects of finite thickness of one dielectric layer that pertains to the top gating of graphene through a high-$\kappa$ dielectric, as well as the effects of finite gap(s) of air between graphene and the nearby dielectric(s).

For graphene laying on a semi-infinite substrate with zero gap, the effects of finite distance of impurities and finite cluster size both give rise to a slightly super-linear dependence of conductivity $\sigma$ on the average charge carrier density $\barn$ in a heavily doped graphene. Taking advantage of the HD model that allows studying 2D structures of impurities with relatively large packing fractions, it is shown that increasing the correlation distance among the impurities gives rise to a strongly increasing slope of $\sigma$ at low $\barn$ values, accompanied by a pronounced sub-linear dependence of conductivity on charge carrier density at higher $\barn$ values. Making reasonable choices of both the impurity density and the correlation distance in the HD model gives good agreement with the experimental data that exhibit sub-linear behavior of the conductivity in graphene,\cite{Tan_2007} whereas inclusion of a perpendicular dipole moment with sufficiently large polarizability also describes the electron-hole asymmetry in soma data.

Reducing the thickness of a high-$\kappa$ dielectric gives rise to an increase in conductivity of graphene at all charge carrier densities in the presence of a 2D distribution of charged impurities and, in particular, causes the conductivity at $\barn=0$ to take finite values. The same conclusions are also true for a homogeneous 3D distribution of impurities throughout the dielectric at low charge carrier densities, but the trend is reversed at higher charge carrier densities because of the pronounced super-linear dependence of the conductivity on $\barn$ as the thickness of the dielectric increases. Further examination of the effects of the dielectric thickness on graphene's mobility, $\mu=\sigma/(e\barn)$, reveals that the existence of a finite gap between graphene and the nearby dielectric and the precise location of a 2D system of impurities both play important roles in the dependence of $\mu$ on charge carrier density. While the role of the distance of the impurities from graphene was discussed before, our results point to the need of including the size of the graphene-substrate gap as another important parameter in modeling the conductivity of graphene.

While the effects of the gap size are also important in the variance of the electrostatic potential in graphene and in the resulting residual charge carrier density within the SCT theory, such effects are seen to gradually diminish in the corresponding conductance minimum as the correlation distance among the impurities in a 2D structure is reduced. This partially justifies the neglect of the graphene-substrate gap in previous studies of the conductivity minimum in the presence of uncorrelated impurities. Finally, reducing the thickness of the high-$\kappa$ dielectric in a top-gated graphene is shown to reduce both the variance of the potential and the resulting residual charge carrier density in graphene, showing that the effects of a system of electron-hole puddles on conductivity in a nominally neutral graphene are likely to be washed-out due to strong screening by a nearby metallic top gate. However, the minimum conductivity would continue to increase with decreasing thickness of the high-$\kappa$ dielectric due to the effect of its geometric capacitance. These opposing roles of the electron-hole puddles in neutral graphene and the geometric capacitance of a dielectric layer in the minimum conductivity of top-gated graphene are worth further exploration.

Summarizing our main findings, we have shown that the effects of finite distance of impurities from graphene, the size of the disk-like clusters of impurities, and the 3D distribution of impurities throughout a dielectric of finite thickness all give rise to super-linear dependence of conductivity on charge carrier density in heavily doped graphene. Next, the thickness of a dielectric and its gap to graphene play important roles in both the conductivity of doped graphene and the conductivity minimum in neutral graphene. Those effects are conveniently taken into account using the electrostatic Green's function for a layered structure of dielectrics. Finally, a strong increase in the slope of conductivity for low charge carrier densities and its saturation at high densities are both well described by large correlation distances among charged impurities in a 2D structure, which may be conveniently described by means of a HD model that allows the use of much higher packing fractions than the simple model of a step-like correlation.

\end{section}

\begin{acknowledgments}
This work was supported by the Natural Sciences and Engineering Research Council of Canada.
\end{acknowledgments}

\appendix

\section{Green's function}

Assume that a single layer of graphene with large area is placed in the plane $z=z_g$ of a Cartesian coordinate system with coordinates $\bR\equiv\{\br,z\}$,
where $\br\equiv\{x,y\}$, and is embedded in a structure that consists of several dielectric layers parallel to graphene, as shown in Fig.~1.
By invoking a translational invariance in the directions of the 2D vector $\br$, one may obtain Green's function (GF) $G(\bR,\bR';t-t')\equiv G(\br-\br';z,z';t-t')$
for the Poisson equation for the entire structure by means of a Fourier transform (FT) with respect to position ($\br\rightarrow\bq$) and time ($t\rightarrow\omega$),
defined via
\begin{eqnarray}
G(\br-\br';z,z';t-t')&=&\int\frac{d^2\bq}{(2\pi)^2}\int\limits_{-\infty}^\infty\frac{d\omega}{2\pi}\,\mbox{e}^{i\bq\cdot(\br-\br')-i\omega(t-t')}\,
\nonumber \\
&&\times
\FG(\bq;z,z';\omega).
 \label{Green}
\end{eqnarray}
If one assumes that graphene has zero thickness, then the FT of the above GF may be expressed in terms of FT of the GF (FTGF) for the dielectric structure \emph{without} graphene, $G^{(0)}(\bR,\bR';t-t')$, as
\begin{eqnarray}
\FG(\bq;z,z';\omega)&=&\FGo(\bq;z,z')
\nonumber \\
&-&
\frac{e^2\chi(q,\omega)\FGo(\bq;z,z_g)\FGo(\bq;z_g,z')}{1+e^2\chi(q,\omega)\FGo(\bq;z_g,z_g)},
 \label{Dyson}
\end{eqnarray}
where $\chi(q,\omega)$ is a 2D, in-plane polarization function of graphene.
We note that this result is easily obtained from a Dyson-Schwinger equation for the full Green's function $\FG(\bq;z,z';\omega)$, which may be generalized to solving a simple matrix algebraic problem for a system of a finite number of graphene layers of zero-thickness that are embedded in a stratified structure of dielectric
slabs described by the FTGF $\FGo(\bq;z,z')$.\cite{Miskovic_2012}

In order to find $\FGo(\bq;z,z')$, we assume that the dielectric structure consists of three layers that occupy the intervals along the $z$ axis defined by $I_1=[-L,0]$, $I_2=[0,H]$ and $I_3=[H,\infty)$, and are characterized by the relative bulk dielectric constants $\ve_j$ with $j=1,2,3$, as shown in Fig.~1. To describe a specific physical configuration,
one may assume that, e.g., the interval $I_1$ is occupied by a high-$\kappa$ dielectric such as HfO$_2$ ($\ve_1\approx 22$) of finite thickness $L>0$, the interval $I_2$ represents a layer of vacuum or air ($\ve_2=1$) of thickness $H\ge 0$ that contains graphene ($z_g\in I_2$), and $I_3$ is a thick (semi-infinite) layer of SiO$_2$ ($\ve_3\approx3.9$). Thus, for finite $z_g>0$ and $H>z_g$, such a configuration allows for finite vacuum gaps of thicknesses $z_g$ and $H-z_g$ between graphene and the dielectrics occupying the intervals $I_1$ and $I_3$, respectively.

The FTGF for the above configuration of dielectric layers, $\FGo(\bq;z,z')$, may be obtained as a tensor $\FGo_{jk}(\bq;z,z')$, where indices $j$ and $k$ correspond to specific locations of the observation point, $z\in I_j$, and the source point, $z'\in I_k$, by solving the FT of the Poisson equation
\begin{eqnarray}
\frac{\partial^2 }{\partial z^2}\FGo_{jk}(z,z')-q^2\FGo_{jk}(z,z')=-\frac{4\pi}{\ve_j}\,\delta_{jk}\,\delta(z-z'),
\label{Poisson}
\end{eqnarray}
where $\delta_{jk}$ is a Kronecker delta with $j,k=1,2,3$, and where we dropped $\bq$ in $\FGo(\bq;z,z')$ for the sake of brevity. When the potential distribution in the system is determined by the potentials at external, ideally conducting electrodes, solutions of Eq.~(\ref{Poisson}) need to satisfy homogeneous boundary conditions of the Dirichlet type at $z=-L$ and $z\rightarrow\infty$ giving
\begin{eqnarray}
\FGo_{1k}(-L,z')&=&0,
 \label{FG_bg}
\\
\FGo_{3k}(\infty,z')&=&0
 \label{FG_tg}
\end{eqnarray}
for $k=1,2,3$. When both $z$ and $z'$ are in the interval $I_j$, one usually defines two components of the corresponding diagonal element of the FTGF as
\begin{eqnarray}
\FGo_{jj}(z,z')=\left\{\begin{array}{ll} \FG_{j}^<(z,z'), \quad & z\le z',
\\
\FG_{j}^>(z,z'), \quad & z'\le z,
\end{array} \right.
\label{FG_1><}
\end{eqnarray}
which must satisfy the continuity and the jump conditions at $z=z'$,
\begin{eqnarray}
\FG_{j}^<(z',z')&=&\FG_{j}^>(z',z'),
 \label{FG_1_cont}
 \\
\left.\frac{\partial}{\partial z}\FG_{j}^>(z,z')\right\vert_{z=z'}-\left.\frac{\partial}{\partial
z}\FG_{j}^<(z,z')\right\vert_{z=z'}&=&-\frac{4\pi}{\ve_j}. \label{FG_1_jump}
\end{eqnarray}
Moreover, assuming abrupt interfaces among various dielectrics, the solution of Eq.~(\ref{Poisson}) needs to satisfy the usual matching conditions at the interfaces $z=0$ and $z=H$ between dielectric regions,
\begin{eqnarray}
\FGo_{1k}(0,z')&=&\FGo_{2k}(0,z'),
 \label{FG_0_cont}
 \\
\ve_1 \left.\frac{\partial}{\partial z}\FGo_{1k}(z,z')\right\vert_{z=0}&=&\ve_2\left.\frac{\partial}{\partial z}\FGo_{2k}(z,z')\right\vert_{z=0},
 \label{FG_0_jump}
\\
\FGo_{2k}(H,z')&=&\FGo_{3k}(H,z'),
 \label{FG_h_cont}
 \\
\ve_2 \left.\frac{\partial}{\partial z}\FGo_{2k}(z,z')\right\vert_{z=H}&=&\ve_3\left.\frac{\partial}{\partial z}\FGo_{3k}(z,z')\right\vert_{z=H},
 \label{FG_h_jump}
\end{eqnarray}
for $k=1,2,3$.

For the sake of definiteness, we assume that charged impurities may only occupy the intervals $I_1$ and $I_2$, so that we only need the elements $\FGo_{jk}$ of the FTGF with $k=1,2$. By solving Eq.~(\ref{Poisson}) subject to the conditions in Eqs.~(\ref{FG_bg})-(\ref{FG_tg}) and Eqs.~(\ref{FG_0_cont})-(\ref{FG_h_jump}), we obtain for $z'\in I_1$\cite{Miskovic_2012}
\begin{eqnarray}
\FGo_{11}(z,z')&=&\frac{4\pi}{\ve_1 q}\,\frac{\sinh\left[q(z_<+L)\right]}{\sinh\left(qL)\right)}
\nonumber \\
&&\times
\frac{\displaystyle{\frac{\ve_1}{\ve_2}}\cosh(qz_>)-\Gamma\sinh(qz_>)}
{\Lambda+\Gamma},
 \label{FGo_11}
\end{eqnarray}
where $z_<=\mathrm{min}(z,z')$, $z_>=\mathrm{max}(z,z')$, $\Lambda\equiv\left(\ve_1/\ve_2\right)\coth(qL)$, and
\begin{eqnarray}
\Gamma=\frac{\ve_2\tanh(qH)+\ve_3}{\ve_2+\ve_3\tanh(qH)},
 \label{Gamma}
\end{eqnarray}
giving
\begin{eqnarray}
\FGo_{21}(z,z')&=&\FGo_{11}(0,z')\left[\cosh(qz)-\Gamma\sinh(qz)\right],
 \label{FGo_21}
\end{eqnarray}
whereas for $z'\in I_2$ we find\cite{Ong_2012}
\begin{widetext}
\begin{eqnarray}
\FGo_{22}(z,z')&=&\frac{\frac{2\pi}{\ve_2 q}}{\Lambda+\Gamma}
\,\left\{\left(\Lambda+\Gamma\right)\mbox{e}^{-q|z-z'|}+\left(\Lambda-1\right)\left(\Gamma-1\right)\cosh\left[q\left(z-z'\right)\right]\right.
\nonumber\\
&&\left.-\left(\Lambda\Gamma-1\right)\cosh\left[q\left(z+z'\right)\right]+\left(\Lambda-\Gamma\right)\sinh\left[q\left(z+z'\right)\right]\right\}.
 \label{FGo_22}
\end{eqnarray}
It is worthwhile mentioning that, with graphene placed at $z_g\in I_2$, one obtains from Eq.~(\ref{FGo_22}) an explicit expression for the background dielectric function $\ebg(q)\equiv 2\pi/\left[q\FGo_{22}(q;z_g,z_g)\right]$ as
\begin{eqnarray}
\ebg(q)=\frac{\ve_2}{2}\frac{\Lambda+\Gamma}{\cosh^2\left(qz_g\right)-\Lambda\Gamma\sinh^2\left(qz_g\right)+
\left(\Lambda-\Gamma\right)\cosh\left(qz_g\right)\sinh\left(qz_g\right)}.
\label{ebg}
\end{eqnarray}
\end{widetext}

For the sake of completeness, we briefly comment on other elements of the FTGF.
One may verify that the symmetry relation $\FGo_{12}(z,z')=\FGo_{21}(z',z)$ is satisfied by defining
\begin{eqnarray}
\FGo_{12}(z,z')=\FG_{22}(0,z')\,\frac{\sinh\left[q(z+L)\right]}{\sinh\left(qL)\right)}.
 \label{FGo_12}
\end{eqnarray}
Moreover, fluctuations of the potential in the interval $I_3$ may be found from
\begin{eqnarray}
\FGo_{3k}(z,z')&=&\FGo_{2k}(H,z')\mathrm{e}^{-q(z-H)},
 \label{FGo_3k}
\end{eqnarray}
with $k=1,2$, which may also be used to deduce components of the FTGF for the source point $z'\in I_3$ via symmetry relations $\FGo_{13}(z,z')=\FGo_{31}(z',z)$ and $\FGo_{23}(z,z')=\FGo_{32}(z',z)$.

Finally, it may be of interest to quote the results for the background dielectric function $\ebg(q)$ and the profile function $\psi(q,z)$ in Eq.~(\ref{Fphizero})
for a few cases of special interest. First, we consider the familiar case of a semi-infinite substrate ($L\rightarrow\infty$) with dielectric constant $\ve_1\equiv\ve_s$ that occupies the region $z<0$, whereas we let $H\rightarrow\infty$ to represent a semi-infinite region $z>0$ of air or vacuum with $\ve_2=1$ that contains a single layer of graphene a distance $z_g\ge 0$ above the substrate. We then obtain
\begin{eqnarray}
\ebg(q)=\left[1-\frac{\ve_s-1}{\ve_s+1}\exp\!\left(-2qz_g\right)\right]^{-1},
 \label{ebg_inf_inf}
\end{eqnarray}
and
\begin{eqnarray}
\psi(q,z)=\left\{\begin{array}{lll}
\displaystyle{\frac{\exp(qz)}{\cosh(qz_g)+\ve_s\sinh(qz_g)}}, \quad & z\le 0,
\\
\\
\displaystyle{\frac{\cosh(qz)+\ve_s\sinh(qz)}{\cosh(qz_g)+\ve_s\sinh(qz_g)}}, \quad & 0\le z\le z_g,
\\
\\
\exp\!\left[-q(z-z_g)\right], \quad & z\ge z_g.
\end{array} \right.
\label{psi_inf_inf}
\end{eqnarray}

As a second example, we consider a semi-infinite substrate ($L\rightarrow\infty$) with dielectric constant $\ve_1$ that occupies the region $z<0$, but we retain $H$ finite and allow for three different dielectric constants as in the original model, and we place graphene at $z_g=H$, i.e., at the boundary between the regions with dielectric constants $\ve_2$ and $\ve_3$. Assuming that the impurities may only reside in the region $z<0$, this configuration describes a case with a dielectric spacer of thickness $H$ between graphene and the region with impurities, giving
\begin{eqnarray}
\ebg(q)=\frac{\ve_3-\ve_2}{2}+\ve_2\left[1+\frac{\ve_2-\ve_1}{\ve_2+\ve_1}\exp\!\left(-2qH\right)\right]^{-1},
 \label{ebg_inf}
\end{eqnarray}
and $\psi(q,z)=\psi_0(q)\,\mbox{e}^{qz}$ for $z< 0$, where
\begin{eqnarray}
\psi_0(q)=
\displaystyle{\frac{\ve_2}{\ve_2\cosh(qH)+\ve_1\sinh(qH)}}.
\label{psi_inf}
\end{eqnarray}

\section{Geometric structure models}

We summarize expressions that define the structure factor for the Hard disk (HD) model due to Rosenfeld\cite{Rosenfeld_1990} for a 2D planar distribution of charged impurities with the packing fraction $p=\pi\nimp r_c^2/4$, where $\nimp=N/\cA$ is their areal number density and $r_c$ is the disk diameter,
\begin{eqnarray}
\SHD(q)&=& \left\{1+16a\left[\frac{J_1(qr_c/2)}{qr_c}\right]^2\right.
\nonumber\\
&+&\left.8b\frac{J_0(qr_c/2)J_1(qr_c/2)}{qr_c}+\frac{8p}{1-p}\frac{J_1(qr_c)}{qr_c}\right\}^{-1}
\label{SHD}
\end{eqnarray}
with
\begin{eqnarray}
a &=& 1+x(2p-1)+\frac{2p}{1-p},
\nonumber\\
b &=& x(1-p)-1-\frac{3p}{1-p},
\nonumber\\
x &=& \frac{1+p}{(1-p)^3}.
\nonumber
\end{eqnarray}
Note that the important long wavelength limit is given by $\SHD(0)=1/x=(1-p)^3/(1+p)$.
The expression in Eq.~(\ref{SHD}) should be compared with the structure factor for a model with the step-like pair correlation function,\cite{Yan_2011,Li_2011}
\begin{eqnarray}
\SSC(q)=1-\frac{8p}{qr_c}J_1(qr_c),
\label{SSC}
\end{eqnarray}
which gives $\SSC(0)=1-4p$.

Next consider a 3D distribution of $N$ point charges $Ze$ occupying the region $-L\le z\le 0$ with a large but finite thickness $L$ and the dielectric constant $\ve_1$, while graphene sits in a region with the dielectric constant $\ve_2$ at the distance $z_g=H\ge 0$. If one disregards the effects of the proximity of graphene and uses the pair correlation (or radial distribution) function for the bulk of a homogeneous charge distribution, $g_{3D}(\br_2-\br_1;z_2-z_1)=g_{3D}(R)$ with $R=\sqrt{(\br_2-\br_1)^2+(z_2-z_1)^2}$, Eqs.~(\ref{cS3D}) and (\ref{psi_inf}) give
\begin{eqnarray}
\cS(q)=\frac{Z^2}{\pi L}\psi_0^2(q)\int\limits_q^\infty\frac{dQ}{Q}\,\frac{S_{3D}(Q)}{\sqrt{Q^2-q^2}},
\label{cS3D_inf}
\end{eqnarray}
where
\begin{eqnarray}
S_{3D}(Q)= 1+N_\mathrm{imp}\int d^3\bR\,\mathrm{e}^{i\bQ\cdot\bR}\left[g_{3D}(R)-1\right],
 \label{S3D_OCP}
\end{eqnarray}
with $N_\mathrm{imp}=N/\left(AL\right)$ being the volume density of particles and $\bQ=\left(\bq,q_z\right)$ a 3D wavevector. For example, we may consider a model for electrostatic correlations among mobile charges in a one-component plasma (OCP)\cite{Ichimaru_1982} at temperature $T$ with the square of the inverse Debye length defined by
$Q_D^2=3\pi N_\mathrm{imp}Z^2e^2/\left(\ve_1 k_B T\right)$, and use the long wavelength result for this system $S_{3D}(Q)=Q^2/\left(Q^2+Q_D^2\right)$ in Eq.~(\ref{cS3D_inf})
to obtain
\begin{eqnarray}
\cS(q)=\frac{Z^2\,\psi_0^2(q)}{2 L\sqrt{q^2+Q_D^2}}.
\label{cS_OCP}
\end{eqnarray}
This result is not used in this work, but it may be found useful in future modeling of the interaction of graphene with an OCP
with a spacer layer of thickness $H$ and dielectric constant $\ve_2$ between graphene and the OCP.


\begin{thebibliography}{10}

\bibitem{Avouris_2012}
P. Avouris and F. Xia, MRS Bulletin \textbf{37}, 1225 (2012).
%Graphene applications in electronics and photonics

\bibitem{Allen_2010} M. J. Allen, V. C. Tung, and R. B. Kaner, Chem. Rev. \textbf{110},132 (2010).
%Honeycomb Carbon: A Review of Graphene

\bibitem{Newaz_2012}
A. K. M. Newaz, Y. S. Puzyrev, B. Wang, S. T. Pantelides, and K. I. Bolotin,  Nat. Commun. \textbf{3}, 734 (2012).
%Probing charge scattering mechanisms in suspended graphene by varying its dielectric environment

\bibitem{Chen_2008}
J. H. Chen, C. Jang, S. Adam, M. S. Fuhrer, E. D. Williams, and M. Ishigami, Nat. Phys. \textbf{4}, 377 (2008).
%Charged Impurity Scattering in Graphene

\bibitem{Tan_2007}
Y.-W. Tan, Y. Zhang, K. Bolotin, Y. Zhao, S. Adam, E. H. Hwang, S. Das Sarma, H. L. Stormer, and P. Kim, Phys. Rev. Lett. 99, 246803 (2007).
%Measurement of Scattering Rate and Minimum Conductivity in Graphene

\bibitem{Fallahazad_2012}
B. Fallahazad, K. Lee, G. Lian, S. Kim, C. M. Corbet, D. A. Ferrer, L. Colombo, and E. Tutuc, Appl. Phys. Lett. \textbf{100}, 093112 (2012),
%Scaling of Al2O3 dielectric for graphene %field-effect transistors

\bibitem{Hollander_2011}
M. J. Hollander, M. LaBella, Z. R. Hughes, M. Zhu, K. A. Trumbull, R. Cavalero,  D. W. Snyder, X. Wang, E. Hwang, S. Datta, and J. A. Robinson, Nano Lett. \textbf{11}, 3601 (2011).

\bibitem{Sarma_2011}
S. Das Sarma, S. Adam, E. H. Hwang, and E. Rossi, Rev. Mod. Phys. \textbf{83}, 407 (2011).

\bibitem{PNAS_2007}
S. Adam, E. H. Hwang, V. M. Galitskii, and S. Das Sarma, Proc. Natl. Acad. USA \textbf{104}, 18392 (2007).

\bibitem{Yan_2011}
J. Yan and M. S. Fuhrer, Phys. Rev. Lett. \textbf{107}, 206601 (2011).

\bibitem{Li_2011}
Q. Li, E. H. Hwang, E. Rossi, and S. Das Sarma, Phys. Rev. Lett. \textbf{107}, 156601 (2011).

\bibitem{McCreary_2010}
K. M. McCreary, K. Pi, A. G. Swartz, W. Han, W. Bao, C. N. Lau, F. Guinea, M. I. Katsnelson, and R. K. Kawakami, Phys. Rev. B \textbf{81},
115453 (2010).

\bibitem{Wehling_2009}
T. O. Wehling, M. I. Katsnelson, and A. I. Lichtenstein, Chem. Phys. Lett. \textbf{476}, 125 (2009).

\bibitem{Ishigami_2007}
M. Ishigami, J. H. Chen, W. G. Cullen, M. S. Fuhrer, and E. D. Williams, Nano Lett. \textbf{7}, 1643 (2007).

\bibitem{Ong_2012}
Z.-Y. Ong and M. V. Fischetti,  Phys. Rev. B \textbf{86}, 121409(R) (2012). %Charged impurity scattering in top-gated graphene nanostructures

\bibitem{Chen_2009}
F. Chen, J. Xia, and N. Tao, Nano Lett. \textbf{9}, 1621 (2009).
%Ionic Screening of Charged-Impurity Scattering in Graphene

\bibitem{Miskovic_2012}
Z.L. Miskovic, P. Sharma and F. O. Goodman, Phys. Rev. B \textbf{86}, 115437 (2012).
%Ionic screening of charged impurities in electrolytically gated %graphene

\bibitem{Castro_2009}
A. H. Castro Neto, F. Guinea, N. M. Peres, K. S. Novoselov, and A. K. Geim, Rev. Mod. Phys. \textbf{81}, 109 (2009).

\bibitem{Gerlach_1986}
E. Gerlach, J. Phys. C: Solid State Phys. \textbf{19}, 4585  (1986).
%Carrier scattering and transport in semiconductors treated by the energy-loss method

\bibitem{Allison_2009}
K. F. Allison, D. Borka, I. Radovic, Lj. Hadzievski, and Z. L. Miskovic, Phys. Rev. B \textbf{80}, 195405 (2009).

\bibitem{Allison_2010}
K. F. Allison and Z. L. Miskovic, Nanotechnology \textbf{21}, 134017 (2010).

\bibitem{Radovic_2012}
I. Radovic, D. Borka, and Z. L. Miskovic, Phys. Rev. B \textbf{86}, 125442 (2012). %Dynamic polarization of graphene by external correlated charges

\bibitem{Krim_2012}
J. Krim, Adv. Phys. \textbf{61}, 155 (2012).
%Friction and energy dissipation mechanisms in adsorbed molecules and molecularly thin films

\bibitem{Newaz_2012_b}
A. K. M. Newaz, D. A. Markov, D. Prasai, and K. I. Bolotin, Nano Lett. \textbf{12}, 2931 (2012).
%Graphene Transistor as a Probe for Streaming Potential

\bibitem{Rosenfeld_1990}
Y. Rosenfeld,  Phys. Rev. A \textbf{42}, 5978 (1990).
%Free-energy model for the inhomogeneous hard-sphere fluid in D dimensions:
%Structure factors for the hard-disk (D =2) mixtures in simple explicit form

\bibitem{Song_2005}
Y.-H. Song, Y.-N. Wang, and Z. L. Miskovic, Phys. Rev. A \textbf{72}, 012903 (2005).

\bibitem{Kaser_1995}
A. Kaser and E. Gerlach, Z. Phys. B \textbf{98}, 207 (1995).
%Scattering of conduction electrons by interface roughness in semiconductor %heterostructures

\bibitem{Kaser_1997}
A. Kaser and E. Gerlach, Z. Phys. B \textbf{103}, 85 (1997).
%Polarizable scattering centers in a fluctuation approach to charge transport

\bibitem{Persson_1991}
B.N. J. Persson, Phys. Rev. B  \textbf{44}, 327 (1991).
%Surface resistivity and vibrational damping in adsorbed layers

\bibitem{Jin_1988}
Z.-C. Jin, L Hamberg, and C. G. Granqvist, J. Appl. Phys. \textbf{64}, 5117 (1988).
%Optical properties of sputter - deposited ZnO:Al thin films

\bibitem{Mendoza_2013}
M. Mendoza, H. J. Herrmann, and S. Succi, Sci. Rep. \textbf{3}, 1052 (2013).
%DOI: 10.1038/srep01052, Hydrodynamic Model for Conductivity in Graphene

\bibitem{Wunsch_2006}
B. Wunsch, T. Stauber, F. Sols, and F. Guinea, New J. Phys. \textbf{8}, 318 (2006).

\bibitem{Hwang_2007}
E. H. Hwang and S. Das Sarma, Phys. Rev. B \textbf{75}, 205418 (2007).

\bibitem{Mowbray_2010}
D. J. Mowbray, S. Segui, J. Gervasoni, Z. L. Miskovic, and N. R. Arista, Phys. Rev. B \textbf{82}, 035405 (2010).

\bibitem{Hansen_1986}
J.-P. Hansen and I. McDonald, \emph{Theory of Simple Liquids}, (Academic, London, 1986).

\bibitem{Mak_2006}
C. H. Mak, Phys. Rev. E \textbf{73}, 065104(R) (2006).
%Large-scale simulations of the two-dimensional melting of hard disks

\bibitem{Guoa_2006}
X. Guoa and U. Riebel, J. Chem. Phys. \textbf{125}, 144504 (2006).
%Theoretical direct correlation function for two-dimensional fluids
%of monodisperse hard spheres

\bibitem{Katsnelson_2010}
M. I. Katsnelson, F. Guinea, and A. K. Geim, Phys. Rev. B \textbf{79}, 195426 (2009).
%Scattering of electrons in graphene by clusters of impurities

\bibitem{Ferreira_2011}
A. Ferreira, J. Viana-Gomes, J. Nilsson, E. R. Mucciolo, N. M. R. Peres, and A. H. Castro Neto, Phys. Rev. B 83, 165402 (2011).
%Unified description of the dc conductivity of monolayer and bilayer graphene at finite densities based on resonant scatterers


\bibitem{Ichimaru_1982}
S. Ichimaru, Rev. Mod. Phys. \textbf{54}, 1017  (1982).
%Strongly coupled plasmas: high-density classical plasmas and degenerate electron %liquids

\bibitem{Maschhoff_1994}
B. L. Maschhoff and J. P. Cowin, J. Chem. Phys. \textbf{101}, 8138 (1994).
%Corrected electrostatic model for dipoles adsorbed on a metal surface

\end{thebibliography}
\end{document}